\documentclass{article}%

\pdfoutput=1
\usepackage{amssymb}
\usepackage{amsfonts}%
\usepackage{amsmath}%
\usepackage{graphicx}

\begin{document}

\title{On first-order definable operations on relational structures}
\author{Bruno COURCELLE\\Bordeaux University, LABRI, CNRS, Talence, France\\courcell@labri.fr}
\date{May 29 2026}
\maketitle


\bigskip

\textbf{Abstract}

We survey the definitions and main properties of first-order (FO) definable
unary operations on relational structures, called \emph{FO-transductions,} and
of FO-definable binary operations based on disjoint union and Cartesian
product.\ We focus our study on \emph{Backwards Translation Theorems} and
\emph{Splitting Theorems} that permit to express FO\ properties of output
structures in terms of finitely many FO properties of the corresponding input ones.\ 

In the particular cases where the operations are defined by
\emph{quantifier-free} (QF) formulas, the quantifier-heights of the obtained
sentences are no larger than those of the input ones.\ It follows that the
class of finite models of a FO\ sentence is \emph{recognizable} with respect
to the considered QF\ operations.\ Recognizability has interesting algorithmic
properties based on finite automata on terms, for structures having bounded
tree-width or clique-width.\ 

We extend our results to FO sentences constructed with \emph{modulo counting
}existential quantifiers.

\bigskip

\textbf{Introduction}

In the context of finite model theory and logical descriptions of trees,
graphs, partial orders and related relational structures, a
\emph{transduction} $\tau$ is a nondetermistic transformation of relational
structures, $\tau:S\mapsto T,$ where $S$ is an $\mathcal{R}$-structure and $T$
is an $\mathcal{R}^{\prime}$-structure. We will only consider finite
structures, and transductions will be defined in logical languages,
first-order (FO) logic and, for comparison, monadic second-order (MSO) logic.

We denote by $\tau(S)$ the set of structures that a transduction $\tau$ can
produce from $S$.\ In many applications, any two structures in $\tau(S)$ are
isomorphic, and we say then that $\tau$ is \emph{deterministic}. We denote by
$\tau(\mathcal{C})$ the set of structures produced by some structure in
$\mathcal{C}$.\ 

Here are some questions relative to these logically defined transductions.

1.\ A classification of graph classes with respect to the relation
$\mathcal{D}\subseteq\tau(\mathcal{C)}$ for FO-transductions has been done in
\cite{BNOS}.\ It extends that of \cite{BluCou} that considers the more
powerful MSO-transductions in this perspective.

2.\ Another concern is to study $\tau^{-1}(\mathcal{D})$, the set of
relational structures\ having an image belonging to an FO- or MSO-definable
class $\mathcal{D}$, and to ask, in particular, the following questions : is
it FO-definable?\ Is it MSO-definable?\ Is it recognizable with respect to an
algebraic structure of interest on the class of input structures (cf.
\cite{CE}, Chapter 4) ? The corresponding answers have implications for
algorithmic metatheorems and theoretical investigations linking logic, graph
structure and algebras of graphs. Some references for algorithmic applications
are \cite{CE}, Chapter 6, and \cite{FriGro,GKS}.

We will classify transductions according to the following criteria, with
obvious inclusions:

A transduction $\tau$ is \emph{scalar} if $D_{T}\subseteq D_{S}$ whenever
$T\in\tau(S)$.

It is \emph{linearly expanding} if we have $D_{T}\subseteq D_{S}\times\lbrack
k]$ for some fixed integer $k$.

It is \emph{vectorial} if $D_{T}\subseteq D_{S}^{k}$ for some fixed integer
$k$.

We will also allow \emph{set parameters}: $T$ is defined, deterministically,
from $S$ and an $m$-tuple $\overline{A}$ of subsets of $D_{S}$. The
nondeterminism comes from the different possible choices of $\overline{A}$.

There are yet other interesting types of transformations of trees, graphs and
related structures : two examples are the \emph{unfolding} of a finite
directed graph into an infinite tree considered in \cite{CouWal}, and the
\emph{universal covering} of a finite undirected graph considered in
\cite{CouMet}.\ They are not of the above types for obvious cardinality reasons.\ 

\emph{MSO-transductions} are specified by MSO formulas and
\emph{FO-transductions} by FO ones.\ FO-transductions can be of the three
above types and MSO-transductions are only of the first two.\ One of our
objectives is to compare the three types of FO-transductions and their properties.\ 

We also consider binary operations of relational structures: \emph{disjoint
union} and different \emph{products} based on Cartesian products of the
domains, enriched with FO\ defined relations of the output structures.\ 

Two fundamental results are the\emph{\ Backwards} \emph{Translation Theorem}
for transductions and the \emph{Splitting Theorem} for binary
operations.\ They permit to express FO\ properties of output structures in
terms of finitely many FO properties of the corresponding input ones.\ In the
particular cases where the operations are defined by \emph{quantifier-free}
(QF) formulas, the quantifier-heights of the obtained sentences are no larger
than those of the input ones.\ It follows that the class of finite models of a
FO\ sentence is \emph{recognizable} with respect to the considered QF\ defined
operations.\ It is well-known that recognizability has interesting algorithmic
properties, based on finite automata on terms, for structures having bounded
tree-width or clique-width.\ 

Finally, we extend our results to \emph{Counting first-order logic} (CFO)
where FO\ formulas can use generalized quantifiers expressing that the number
of elements satisfying a formula has cardinality $p$ modulo $q$. This
extension has been already consided in \cite{KuSch,NesPOM+}.

\bigskip

The present article is more a survey than a source of new results. Its purpose
is to clarify and classify the definitions and properties of the above
mentioned FO definable operations on relational structures.\ 

\bigskip

\textbf{Overview of sections: }Section 1 consists of definitions.\ Section 2
presents and studies scalar FO-transductions frequently called
\emph{interpretations}.\ Section 3 discusses quotient structures, defined by
particular scalar transductions.\ Section 4 defines linearly expanding
FO-transductions as restrictions of MSO-transductions. Section 5 discusses
union and products of disjoint relational structures and Section 6\ vectorial
FO-transductions.\ Section 7 is devoted to recognizability and Section 8 to
the extensions of the preceding results to CFO-transductions.

\section{Definitions}

All graphs and relational structures will be finite. The cardinality of a
finite set $A$ is denoted by $\left\vert A\right\vert .$ We denote by $A\uplus
B$\ the union of two disjoint sets $A$ and $B$, and by $\biguplus$ its
extension to several sets. We denote by $[k]$ the set $\{1,...,k\}$ and by
$[0,k]$ the set $\{0,...,k\}.$

We write $\mathbb{C}^{(p,q)}(A)\ $\ to express that a set $A$\ has cardinality
$p$ modulo $q$.

\bigskip

\textbf{Definitions 1.1} : \emph{Relational structures and FO\ formulas}

(a) A \emph{signature} is a finite set $\mathcal{R}$ of relations, having each
a fixed arity.\ An $\mathcal{R}$-\emph{structure} is $S=(D_{S},R_{S},...)$
where $D_{S}$ is its finite (possibly empty) \emph{domain} and each $R_{S}%
$\ is a relation on $D_{S}$\ of correct arity.

As in \cite{CE}, we might allow constants in structures. Let $\mathcal{R}_{0}
$ be a finite set of nullary symbols.\ In an ($\mathcal{R}\cup\mathcal{R}_{0}%
$)-structure $S$, the value $a_{S}$ of $a$ in $\mathcal{R}_{0}$ is called a
\emph{source}.\ It is frequently simpler to represent $a_{S}$ by the unary
relation $Eq_{a}$, such that $Eq_{a}(x)$ holds if and only if $x=a_{S}$.\ 

Isomorphism of structures (graphs, trees, etc.) is denoted by $\simeq$. A
\emph{class} of structures is a set of structures closed under isomorphism. We
consider isomorphic structures as essentially identical, however concrete
structures, defined as above, are necessary for constructions.

The union $S\oplus T$ of two disjoint $\mathcal{R}$-structures $S$ and
$T$\ has domain $D_{S}\uplus D_{T}$\ and relations defined similarly, as
unions of those of $S$ and $T$.\ Each tuple of $R_{S\oplus T}$\ is either in
$D_{S}$\ or in $D_{T}$. If $S$ and $T$ are not disjoint, then on replaces one
by an isomorphic disjoint copy.\ Hence we can define $S\oplus S$, up to
isomorphism, and even concretely as we will see in Definition 4.1.

(b) \emph{First-order} (FO) formulas over $\mathcal{R}$ are written with = and
the relations of $\mathcal{R}$, they express properties of such structures. A
\emph{sentence} is a formula without free variables.

\emph{Quantifier-free} (QF)\ formulas are those without quantifiers. As we do
not allow constants, QF sentences are trivial, either $True$ or $False$.

(c) For denoting subsets of the considered domains, we allow set variables $X
$ in FO formulas and the atomic formulas $x\in X$.\ We exclude quantifications
over set variables. The notation $\varphi(x_{1},...x_{n},X_{1},...,X_{m})$
means that the free variables of $\varphi$ are among $x_{1},...,x_{n}$,
$X_{1},...,X_{m}.$

We will abreviate a tuple of individual variables $(x_{1},...,x_{k})$ by
$\overline{x}$ and a corresponding $k$-tuple of values in the domain of a
logical structure by $\overline{a}$. For $X_{1},...,X_{m}$ we will use
similarly $\overline{X}$ and $\overline{A}$.

A formula $\varphi(\overline{x},\overline{X})$ expresses a property of
$(S,\overline{a},\overline{A})$, that is, of an $\mathcal{R}$-structure $S$
enriched with distinguished elements (in $\overline{a}$) and subsets (in
$\overline{A}$) of the domain, that we can consider as \emph{colors}.

Actually, $(S,\overline{A})$ can be consisered as an ($\mathcal{R\cup}$
$\overline{X}$)-structure and $\varphi(\overline{x},\overline{X})$ as an
formula of signature $\mathcal{R\cup}\overline{X}$ with free variables in
$\overline{x}$. This remark applies also to MSO\ formulas, where set variables
can be quantified.\ Hence, results concerning formulas without free set
variables extend to formulas with such.

A\ QF\ formula has \emph{quantifier-height }0.\ The quantifier-height of
$\varphi\vee\psi$, $\varphi\wedge\psi$ and $\lnot\varphi$ is the maximum of
those of $\varphi,\psi.\ $Each quantification adds one to the
quantifier-height of the considered formula.

\bigskip

\textbf{Example 1.2} : \emph{Graphs}.

A simple directed graph (without parallel edges) is represented by the
relational structure $(V_{G},edg_{G})$ where $V_{G}$ is the set of vertices
and $edg_{G}(x,y)$ means that there is a directed edge $x\rightarrow y$. In
\cite{CE}, we consider graphs with sources and operations on them.\ See below
Definition 3.4.\ 

\bigskip

\textbf{Definitions 1.3 }: \emph{Transductions.}

(a) A \emph{transduction} $\tau$ of type $\mathcal{R}\rightarrow
\mathcal{R}^{\prime}$ is a transformation of relational structures that maps
an $\mathcal{R}$-structure $S$ to a set $\tau(S)$ of $\mathcal{R}^{\prime}%
$-structures. We require that if $S\simeq S^{\prime}$, then $\tau(S)\simeq
\tau^{\prime}(S^{\prime})$ which means that each $T$\ in $\tau(S)$ is
isomorphic to some $T^{\prime}$ in $\tau(S^{\prime})$ and vice versa. It is
\emph{functional} if each $\tau(S)$ is empty or singleton.\ It is
\emph{semantically deterministic} if any two structures in $\tau(S)$ are
isomorphic. If the set $\tau(S)$ is empty, then we say $\tau$ is undefined for
$S$.

We will consider transductions\ specified by logical formulas, in most cases, first-order.

(b) We denote by $\tau(\mathcal{C})$ the set of structures produced from some
structure in $\mathcal{C}$.\ We denote by $\tau^{-1}(\mathcal{D})$, the set of
$\mathcal{R}$-structures\ having an image in $\mathcal{D}$.\ 

(c) A transduction $\tau$ is \emph{scalar} if $D_{T}\subseteq D_{S}$ whenever
$T\in\tau(S)$.

It is \emph{linearly expanding} if $D_{T}\subseteq D_{S}\times\lbrack k]$ for
some fixed integer $k$,

It is \emph{vectorial} if $D_{T}\subseteq D_{S}^{k}$ for some fixed integer
$k$.

(d) The nondeterminism comes from the use of sets call \emph{parameters}. A
transduction $\tau$ using parameters takes as arguments a structure $S$ and an
$n$-tuple $\overline{A}$ of subsets of its domain given as values of an $n
$-tuple of set variables.\ Different outputs can be obtained from different
tuples $\overline{A}$.\ \hfill$\square$

\bigskip

The monadic second-order (MSO) transductions studied in \cite{CE} are, by
definition, linearly expanding.\ First-order (FO)\ transductions can be
vectorial, as we will see.

Unfolding and universal covering produce infinite trees from finite graphs,
hence are not of the above types. We only consider finite structures.

Continuing the remarks in Definition 1.1(c), we can consider a transduction
$\tau$ with parameters $X_{1},...,X_{m}$ from $\mathcal{R}$-structures to
$\mathcal{R}^{\prime}$-structures as the functional transduction that
transforms the\ $(\mathcal{R\cup}\overline{X})$-structure $(S,\overline{A}%
)$\ where the $m$-tuple $\overline{A}$ defines values for $\overline{X}$.\ We
can thus in many cases restrict proofs to transductions without parameters.

\bigskip

\textbf{Example 1.4 : }The mapping that transforms a graph by eliminating its
isolated vertices and adding the directed edges from $x$ to $y$ whenever there
is a vertex $z$ such that $x\rightarrow z$ and $z\rightarrow y$ is a
functional scalar FO\ transduction.

\bigskip

\textbf{Definition 1.5} : \emph{Counting first-order logic }

We extend FO logic by means of \emph{generalized quantifiers} as follows :

\begin{quote}
$\mathbb{C}^{(p,q)}\overline{x}.\varphi$ means in a relevant logical
structure, that the set of $k$-tuples $\overline{a}$ of values of
$x_{1},...,x_{k}$ (forming $\overline{x}$) satisfying $\varphi$\ has
cardinality $p$ modulo $q$.
\end{quote}

For an example, $\mathbb{C}^{(1,2)}(x,y).edg(x,y)$ expresses that a simple
directed graph has an odd number of edges (cf.\ Example 1.2). In Section 8, we
will extend to Counting (modulo) First-Order logic (CFO in short), our results
on FO\ transductions.

The article\ \cite{KuSch} also considers conditions of the form $\left\vert
A\right\vert \in R$ where $R$ is a recursive set of integers, but this is not
our present purpose.

\section{Scalar FO\ transductions}

Scalar FO transductions are particular MSO\ transductions, which are defined
below in Definition 3.1\ and, with more details, in Chapter 7\ of \cite{CE}.

\bigskip

\textbf{Definition 2.1 }: \emph{Scalar FO transductions.}

(a) A \emph{functional scalar FO-transduction} $\tau$ of type $\mathcal{R}%
\rightarrow\mathcal{R}^{\prime}$ is defined without parameters and transforms
an $\mathcal{R}$-structure $S$ into an $\mathcal{R}^{\prime}$-structure\ $T$
such that$\ D_{T}$ is a subset of $D_{S}$ defined in $S$\ by an FO\ formula
$\delta(x)$ called the \emph{domain formula}.\ (Such a transduction is
frequently called an \emph{interpretation}).\ If $R$ is a $k$-ary relation in
$\mathcal{R}^{\prime}$, then $R_{T}$\ is defined as the set of $k$-tuples
$\overline{a}$ in $D_{S}$\ satisfying a formula of the form $\delta
(x_{1})\wedge\delta(x_{2})\wedge...\wedge\delta(x_{k})\wedge\theta_{R}%
(x_{1},...,x_{k}).$ Furthermore, a sentence $\chi$, called the
\emph{precondition}, expresses whether $\tau(S)$ is defined.\ 

Hence $\tau$ is specified by a tuple of formulas $\Delta:=(\chi,\delta
(x),\theta_{R}(x_{1},...,x_{k}),...,$\ $\theta_{R^{\prime}}(x_{1}%
,\ \ ...,x_{k^{\prime}}))$\ called a \emph{definition scheme}, where
$R,...,R^{\prime}$ are the relations of $\mathcal{R}^{\prime}$. It is an
\emph{FO}\ (resp. a \emph{CFO}) \emph{transduction} if all these formulas are
FO (resp. CFO\footnote{See Section 8 for details.}). It is a
\emph{QF-transduction} if all defining formulas are quantifier-free ($\chi$
must be \emph{true}). Note that $T$ is uniquely defined from $S$. We denote by
$I_{\Delta}$\ the associated transduction.\ 

(b) Here is an extension of this definition: a scalar FO transduction $\tau$
may use \emph{set parameters}.\ It transforms $(S,\overline{A})$ into $T$,
where $\overline{A}$ is an $m$-tuple of subsets of $D_{S}$. The defining
formulas are written with set variables $Y_{1},...,Y_{m}$ intended to
hold\ the parameters of $\overline{A}$.\ In that case, $\chi$ is not a
sentence, it expresses that the parameters are "correctly" chosen. The
parameters need not subsets of $D_{T}$.\ Note that $\tau(S,\overline{A})$ may
be undefined for some values of $\overline{A}$. The precondition $\chi$ may
impose to the parameters to be singletons.

The pair $(S,\overline{A})$ can be considered as an expansion of the structure
$S$\ by $m$ unary relations. Hence, many proofs for the basic case extend
immediately to transductions using set parameters.\ (See however Theorem 4.4
for some technical details.)

Such a transduction can be seen as transforming $S$ into the set of structures
$\tau(S,\overline{A})$, for all acceptable tuples $\overline{A}.$

\bigskip

\textbf{Example 2.2 :} (a) The mapping that transforms a graph by eliminating
its isolated vertices and adding directed edges from $x$ to $y$ whenever there
is a vertrex $z$ such that $x\rightarrow z$ and $z\rightarrow y$ is a scalar
functional FO-transduction.\ It uses no parameters.

(b) The mapping that transforms a graph by eliminating the edges between any
two vertices of a set $A$, is a QF-transduction using a set parameter.

\bigskip

\textbf{Theorem 2.3 (}\emph{Backwards Translation})\ : Let $\tau$ be a scalar
FO-transduction defined by a definition scheme $(\chi,\delta,...)$ without
parameters of type $\mathcal{R}\rightarrow\mathcal{R}^{\prime}$. Let
$\varphi(x_{1},...,x_{k})$ be an FO-formula over $\mathcal{R}^{\prime}$.\ 

(1) One can construct an FO\ formula $\psi(x_{1},...,x_{k})$ over
$\mathcal{R}$\ such that,

\begin{quote}
for each $k$-tuple $\overline{a}$ in $D_{S}^{k}$, we have

$\overline{a}$ $\in$ $D_{T}^{k}$ and $(T,\overline{a})\models\varphi$ (also
denoted by $T\models\varphi(\overline{a})$ ) if and only if

$(S,\overline{a})\models\delta(x_{1})\wedge\delta(x_{2})\wedge...\wedge
\delta(x_{k})\wedge\psi(x_{1},...,x_{k}).$
\end{quote}

(2) If $\tau$ is a QF-transduction, the quantifier-height of $\psi$ is at most
that of $\varphi$.

(3) Let now $\varphi(x_{1},...,x_{k},X_{1},...,X_{m})$ have free set variables
$X_{1},...,X_{m}$. One can construct an FO\ formula $\psi(x_{1},...,x_{k}%
,X_{1},...,X_{m})$ such that,

\begin{quote}
for each $k$-tuple $\overline{a}$ in $D_{S}^{k}$, each $\overline{A}$ in
$\mathcal{P}(D_{S})^{m}$ , we have

$\overline{a}$ $\in$ $D_{T}^{k}$ , $\overline{A}$ $\in\mathcal{P}(D_{T})^{m}$
and $(T,\overline{a},\overline{A})\models\varphi$ if and only if

$(S,\overline{a},\overline{A})\models\delta(x_{1})\wedge...\wedge\delta
(x_{k})\wedge\forall x.(x\in X_{1}\cup...\cup X_{m}\Longrightarrow\delta(x))$

$\qquad\qquad\qquad\wedge\psi(x_{1},...,x_{k}X_{1},...,X_{m}).$ \hfill
$\square$
\end{quote}

\bigskip

Theorem 7.10\ in the book \cite{CE} proves a more general statement concerning
monadic second-order (\emph{MSO-}) transductions.

\bigskip

\textbf{Proof }: (1) The proof is an easy inductive construction.\ We only
give representative examples. If $\varphi$ is $R(x_{i},x_{j}),$ then $\psi$ is
$\theta_{R}(x_{i},x_{j}).$\ If $\varphi$ is $\exists x_{n+1}.\varphi^{\prime
}(x_{1},...,$ $x_{n+1}),$ then $\psi(x_{1},...,x_{k})$\ is taken as $\exists
x_{n+1}.\delta(x_{n+1})\wedge\psi^{\prime}(x_{1},...,x_{n+1}) $\ where
$\psi^{\prime}$\ is built from $\varphi^{\prime}$ by way of induction.

(2) By inspecting the inductive proof, we can see that quantifier-height is
not increased because $\delta$ and the $\theta_{R}$'s are QF.

(3) This extension of (1) is clear.\ If $\varphi$ is QF, then the resulting
formula has quantifier-height 1 because of the clause $\forall x.(x\in
X_{1}\cup...\cup X_{m}\Longrightarrow\delta(x))$.\ $\ $ \hfill$\square$

\bigskip

\textbf{Corollary 2.4 }: (1) If $\tau$ is an FO transduction without
parameters and $\mathcal{D}$ is FO-definable, then so is $\tau^{-1}%
(\mathcal{D})$.\ 

(2) If $\tau$ is an FO transduction with parameters $Y_{1},...,Y_{m}$ and
$\mathcal{D}$ is FO-definable, then $\tau^{-1}(\mathcal{D})$ is $\exists
$MSO-definable, which means that it is defined by a sentence of the form
$\exists Y_{1},...,Y_{m}$.$\varphi$ where $\varphi$\ is FO.

\bigskip

\textbf{Remarks} \textbf{2.5}\ : (1) In Definition 2.1, we could include the
conditions $\delta(x_{1})\wedge...\wedge\delta(x_{k})$ in the formula
$\theta_{R}(x_{1},...,x_{k}).$\ However this would yield some
redundancies.\ Let for example $\theta_{R}(x,y)$\ be $\delta(x)\wedge
\delta(y)\wedge R_{1}(x)\wedge R_{2}(y)$.\ Then the translation of $\exists
x,y.R(x,y)\wedge R(y,x)$ would be

\begin{center}
$\exists x,y.[\delta(x)\wedge\delta(y)\wedge(\delta(x)\wedge\delta(y)\wedge
R(x,y))\wedge(\delta(y)\wedge\delta(x)\wedge R(y,x))].$
\end{center}

(2) If\ a class $\mathcal{C}$ of $\mathcal{R}$-structures is FO-definable and
$\tau$ is an FO-transduction, then $\tau(\mathcal{C})$ need not be
FO-definable.\ Let for example $\mathcal{C}$ be the set of words $a^{n}b^{n}$
defined as paths with vertices alternatively colored by $a$ and $b$.\ Let
$\tau$ delete the edges.\ The resulting structures consist of the finite sets
having as many $a$-labelled elements as $b$-labelled ones.\ It is not
FO-definable (and not MSO-definable either).\ See \cite{CE} Proposition 5.13(1).

\bigskip

The following theorem is a consequence of Theorem 2.3, where the general
FO-transductions may use parameters.

\bigskip

\textbf{Theorem 2.6}\ : (1) The composition of two scalar FO-transductions is
an FO-transduction.

(2) The composition of two scalar QF-transductions without parameters is a
QF-transduction without parameters.

\textbf{Proof: }We first consider the composition of two FO-transductions
without parameters (hence functional ones).\ Let $\tau$ be of type
$\mathcal{R}\rightarrow\mathcal{R}^{\prime}$ and $\tau^{\prime}$ be of type
$\mathcal{R}^{\prime}\rightarrow\mathcal{R}^{\prime\prime}$ defined by
definition schemes $\Delta=(\chi,\delta,\theta_{R},...)$ and\ $\Delta^{\prime
}=(\chi^{\prime},\delta^{\prime},\theta_{R^{\prime}},...)$. Let $\tau
^{\prime\prime}:=\tau^{\prime}\circ\tau$ of type $\mathcal{R}\rightarrow
\mathcal{R}^{\prime\prime}$.

Let $\tau$ transform $S$ into $T$ and $\tau^{\prime}$ transform $T$ into $U$.
Then $\tau^{\prime\prime}(S)$ is defined if and only if $\tau(S)$ is defined,
say $\tau(S)=T$, and so is $\tau^{\prime}(T)$, that is if and only if $S$
satisfies $\chi\wedge\widehat{\chi^{\prime}}$ where $\widehat{\chi^{\prime}}%
$\ is obtained by Theorem 2.3(1) from $\chi^{\prime}$ and $\tau$.\ Hence the
precondition of $\tau^{\prime\prime}$\ is that sentence.

Similarly, an element $x$ in $D_{S}$\ belongs to $D_{U}$ if and only if
$S\models\delta(x)\wedge\widehat{\delta^{\prime}}(x)$ where $\widehat
{\delta^{\prime}}$ is obtained by Theorem 2.3(1) from $\delta^{\prime}$ and
$\tau$, which gives the domain formula $\delta^{\prime\prime}$ of the
definition scheme $\Delta^{\prime\prime}$ intended to specify $\tau
^{\prime\prime}.$

\bigskip

Next we consider a relation $R\in\mathcal{R}^{\prime}$ that we take binary for
simplicity.\ A pair $(x,y)\in D_{S}\times D_{S}$\ belongs to $R_{U}$ if and
only if $S$ satisfies $\delta^{\prime\prime}(x)\wedge\delta^{\prime\prime
}(y)\wedge\widehat{\theta_{R}^{\prime}}(x,y)\ $where $\widehat{\theta
_{R}^{\prime}}$ is obtained by Theorem 2.3(1) from $\widehat{\theta
_{R}^{\prime}}$ and $\tau$. This proves Assertion (1) for FO transductions
without parameters and also Assertion (2).

\bigskip

We now prove (1) for transductions using parameters. Let $\tau$ use parameters
$X_{1},...,X_{n}$ and $\tau^{\prime}$ use $Y_{1},...,Y_{m}$. The transduction
$\tau^{\prime\prime}$\ will use parameters $X_{1},...,X_{n}$, $Y_{1}%
,...,Y_{m}$.

Let $\overline{A}$,$\overline{B}$ be an ($n+m)$-tuple of subsets of $D_{S}%
$.\ We need to define $\tau^{\prime}(T,\overline{B})$ where $T:=\tau
(S,\overline{A})$.\ In particular, the components of $\overline{B}$\ must be
subsets of $D_{T}$.\ We must have:

\begin{quote}
$(S,\overline{A},\overline{B})\models\chi\wedge\widehat{\chi^{\prime}}%
(X_{1},...,X_{n},Y_{1},...,Y_{m})$

$\qquad\qquad\qquad\wedge\forall y.[y\in Y_{1}\cup...\cup Y_{m}\Longrightarrow
\delta(x,X_{1},...,X_{n})]$
\end{quote}

where as above $\widehat{\chi^{\prime}}$ is obtained by Theorem 2.3(3).\ The
remaining parts of $\Delta^{\prime\prime}$ are also constructed by means of
Theorem 2.3(3).\ Note the increase of quantifier-height.$\ $ \hfill$\square$

\section{Quotient structures}

If $S$ is an $\mathcal{R}$-structure and $\equiv$ is an equivalence relation
on $D_{S}$, then $S/\equiv$ is the $\mathcal{R}$-structure whose domain is
$D_{S}/\equiv$\ (the quotient set) and relations such that $R_{S/\equiv
}([d_{1}],...,[d_{k}])$ holds (in $S/\equiv)$ if and only if some
$R(a_{1},...,a_{k})$ holds in $S$ where, for each $i$, the element $a_{i}$ is
equivalent to $d_{i}$. Here $[d]$ denotes the equivalence class of $d$ w.r.t.
$\equiv$.

We now examine the transduction $S\longmapsto T:=S/\equiv$ between
$\mathcal{R}$-structures in the case where $x\equiv y$ is defined in $S$\ by a
formula $\xi(x,y)$.\ Although we focus on FO\ logic, the definitions below
extend to MSO and related logical languages.\ 

For having a scalar transduction, we construct $D_{T}$ as a subset of $D_{S}%
$\ obtained by choosing in each equivalence class $[d],$ a \emph{canonical
representative} $u$, that is (hopefully) FO-definable from $d$.

\bigskip

\textbf{Theorem 3.1}\ : Let $S$ be an $\mathcal{R}$-structure equipped with
an\ FO-definable equivalence relation $\equiv$.

(1) If a canonical representative $u$ of any equivalence class $[d]$ is
FO-definable from $d$, then the mapping $S\longmapsto T\simeq S/\equiv$ is a
scalar FO-transduction. The domain $D_{T}$\ is the set of representatives of
the equivalence classes.

(2) If $\leq$\ is an arbitrary linear order on\ $D_{S}$, then the mapping
$(S,\leq)\longmapsto T\simeq S/\equiv$ is a scalar FO-transduction, where
$D_{T}$ is the set of $\leq-$minimal elements of the equivalence classes.\ Any
two linear orders yield isomorphic structures $T$ under this transduction.

\textbf{Proof }: (1) The domain formula $\delta(u)$ says that $u$ represents
some class $[d]$. The other definitions are straightforward.\ First-order
quantifications are needed.

(2) The next construction helps if we have no way to determine by an
FO\ formula a canonical element in each class.\ If $S$ is linearly ordered by
some extra relation, then the canonical element of a class can be FO-defined
as the minimal one. The result follows by (1).\ $\ $ \hfill$\square$

\bigskip

In Case (2), we get from different choices of $\leq$ different but isomorphic
structures.\ We say that the obtained transduction is \emph{order-invariant}.
See \cite{CouX}\ for order-invariant MSO-transductions.

\bigskip

\textbf{Examples 3.2\ }: (1) Let $S:=(V,edg,R)$ define a graph $G=(V,edg)$
with a distinguish set $R$\ of vertices handled as a unary relation.\ We let
$x\equiv y\ $mean that $x=y$ or $R(x)\wedge R(y)$ holds. Then $S/\equiv$
defines the graph where all vertices in $R$ are fused into a single one, which
may create a loop.\ The mapping $S\longmapsto S/\equiv$ is an order-invariant
FO-transduction.\ Without the order, no FO\ formula can distinguish one among
several vertices of $R$.

(2) Let $S:=(N,\leq,lab_{a},...,lab_{d})$ represent a word $w$ having $N$ as
set of positions, defined by a linear order (as opposed to a successor
relation\footnote{In this case, the corresponding linear order on positions is
not FO-definable.\ See \cite{CouX}.}).\ An element $x$ of $N$ is a position of
letter $a$ if $lab_{a}(x)$ holds. Let $x\equiv y$ mean that $x$ and $y$ are
positions of a same letter, say $u$, and all positions between them are
so.\ Hence $S/\equiv$ is the word obtained from $w$ by fusing any two
positions holding a same letter. The quotient mapping is an
FO-transduction.\ In this case, the order $\leq$ is part of $S$ and is not an
auxiliary order as in Theorem 3.1(2).$\ $ \hfill$\square$

\bigskip

\textbf{Remarks 3.3} : (1) Theorem 3.1\ extends to equivalence relations and
representative elements that are definable in CFO (cf. Definition\ 1.5 and
Section 8) or in MSO\ and its extensions, CMSO and $\leq$-MSO, cf. \cite{CouX}.

(2) In \ \cite{Coudesc} we see cases where $D_{T}$\ is an FO definable
quotient of $D_{S}^{k\ }$: from a linear order on $D_{S}$, one gets an FO
definable lexicographically minimal element of each equivalence
class.\ Examples are in \cite{Coudesc} and in Example 6.2(b) below.

(3) If $S$ is countably infinite, we can apply Theorem 3.1(2) by augmenting
$D_{S}$\ with a linear order whose order type is $\omega,$ cf. \cite{CouX}%
\footnote{We have no motivation for considering uncountable structures.}.

\bigskip

\textbf{Definition 3.4}\ : \emph{Parallel composition of graphs with
sources\ }(\cite{CE}, Definition 2.25)

Let $A$ and $B$\ be two sets of nullary symbols, that we call source names.
Let $G=(V,edg,(a_{G})_{a\in A})$ be a graph with sources, we let
$src(G):=A$.\ We have $a_{G}\neq b_{G}$ if $a\neq b$. Let $H$ be disjoint from
$G$ and similarly, $src(H):=B.$ The $G//H$ is defined as the quotient
$(G\oplus H)/\equiv$ where $x\equiv y$ if and only if $x=y,$ or$\ x=a_{G}$ and
$y=a_{H}$ for some $a$ in $A\cap B,$ or vice-versa. We will examine this
operation in Section 5.1.

The extension to \emph{relational structures with sources} is straightforward.

\section{Linearly expanding transductions}

We will consider transductions that are restrictions to first-order formulas
of the MSO-transductions of \cite{CE}, Section 7.1 (also considered in
\cite{BluCou, CouX, CouWal}). We recall the corresponding definitions.

\subsection{Monadic Second--Order transductions}

We first consider transductions without parameters.\ Parameters can be handled
as unary relations added to the signature $\mathcal{R}$ of input structures,
as explained above in Definition 1.3.

\bigskip

\textbf{Definition 4.1}\ : \emph{MSO-transductions}

(1) Let $\mathcal{R}$ and $\mathcal{R}^{\prime}$ be relational signatures.\ A
$k$-\emph{copying, parameterless} \emph{MSO-transduction} $\tau$ of type
$\mathcal{R}\rightarrow\mathcal{R}^{\prime}$\ transforms an $\mathcal{R}%
$-structure $S$ into an $\mathcal{R}^{\prime}$-structure $T$ such that:

\begin{quote}
$D_{T}:=D_{1}\times\{1\}\cup...\cup D_{k}\times\{k\}\subseteq D_{S}%
\times\lbrack k].$
\end{quote}

(a) It uses a \emph{precondition}, that is, a sentence $\chi$ expressing (by
$S\models\chi$) that $\tau(S)$ is well-defined.

(b) A formula $\delta_{i}(x)$ defines $D_{i}:=\{d\in D_{S}\mid S\models
\delta_{i}(d)\}.$ ($T$ can the empty structure).\ 

(c) An $n$-ary relation $U_{T}$\ for $U$ in $\mathcal{R}^{\prime}$ is defined by

\begin{quote}
$((a_{1},j_{1}),...,(a_{n},j_{n}))\in U_{T}:\Longleftrightarrow$ $\overline
{U}_{j_{1},...,j_{n}}(a_{1},...,a_{n})$ holds
\end{quote}

where\ formulas $\rho_{j_{1},...,j_{n}}(x_{1},...,x_{n})$\ define the
relations $\overline{U}_{j_{1},...,j_{n}}$'s\ (here $j_{1},...,$ \ $j_{n}%
\in\lbrack k]$)\ \ by :

\begin{quote}
$\overline{U}_{j_{1},...,j_{n}}(a_{1},...,a_{n}):\Longleftrightarrow
S\models\delta_{j_{1}}(a_{1})\wedge...\wedge\delta_{j_{n}}(a_{n})\wedge
\rho_{j_{1},...,j_{n}}(a_{1},...,a_{n}).$
\end{quote}

All these formulas are MSO.\ They form a \emph{definition scheme}
$\Delta:=(\chi,\delta_{1},...,\delta_{k},$ \ \ $...,\rho_{j_{1},...,j_{n}%
},...). $

The corresponding transduction is linearly expanding by definition. It is a
partial function from $\mathcal{R}$-structures to $\mathcal{R}^{\prime}%
$-structures because $\tau(S)$ is undefined if $S\models\lnot\chi$.

(2) It may be useful to put in $\mathcal{R}^{\prime}$ the unary relations
$In_{i}$, for $i\in\lbrack k],$ such that we have in $T$:

\begin{quote}
$(a,j)\in In_{i}:\Longleftrightarrow i=j$ and $S\models\delta_{i}(a).$
\end{quote}

They characterize in $T$\ the subsets $D_{i}\times\{i\}.$ One can also put in
$\mathcal{R}^{\prime}$ the binary relation $Clone$ defined such that, for
$i,j\in\lbrack k]$ we have :

\begin{quote}
$(a,i),(b,j)\in Clone:\Longleftrightarrow i\neq j$ and $S\models
a=b\wedge\delta_{i}(a)\wedge\delta_{j}(a).$
\end{quote}

It expresses that two elements of $D_{T}$\ "come from" a same element (here
$a$) of $D_{S}$.

(3) We obtain \emph{linearly expanding FO-transductions} and
\emph{QF-transductions} by restricting the defining formulas to be FO or QF\ respectively.

(4) These transductions may use set parameters as in Definition 2.1,
considered as expanding the input structures. \hfill$\square$

\bigskip

Note that the transduction $S\longmapsto S\oplus S$\ (of Definition 1.1) is a
2-copying QF-transduction without parameters.

\bigskip

\textbf{Theorem 4.2\ }: \emph{Backwards Translation Theorem} (\cite{CE},
Theorem 7.10).

Let $\tau$ be an FO transduction of type $\mathcal{R}\rightarrow
\mathcal{R}^{\prime}$ and $\varphi$ be an FO\ sentence of signature
$\mathcal{R}^{\prime}$.\ 

(1) One can construct an FO\ sentence $\psi$ of signature $\mathcal{R}$\ such
that, for every structure $S$, we have $\tau(S)\models\varphi$ if and only
if$\ S\models\psi.$

(2) If $\tau$ is a QF-transduction, the quantifier height of $\psi$ is at most
that of $\varphi$.$\ $ \hfill$\square$

\bigskip

For proving (2), it is enough to observe that in the proof given in \cite{CE},
no set quantification is introduced to construct $\psi$. The slightly more
complicated case where $\varphi$ has free variables $x_{1},...,x_{n},X_{1},$
\ $...,X_{p}$ is considered in \cite{CE}, Theorem 7.10.

\ 

\textbf{Definition 4.3} : \emph{Transductions with parameters}.

A \emph{transduction with parameters} of type $\mathcal{R}\rightarrow
\mathcal{R}^{\prime}$\ transforms an $\mathcal{R}$-structure $S$ given with an
$m$-tuple $\overline{A}$ of subsets of $D_{S}$ (called the \emph{parameters}%
)\ into an $\mathcal{R}^{\prime}$-structure $T=\tau(S,\overline{A})$ where
$\tau$ is as in Definition 4.1.\ Note that $\tau(S,\overline{A}) $ is
undefined if the precondition $\chi(\overline{A})$ is not satisfied in $S$.
Here, $\chi$ has free set variables intended to hold the parameters.

We consider $\tau$ as an nondeterministic transduction transforming $S$\ into
the set $\tau(S)$ of (well-defined) $\mathcal{R}^{\prime}$-structures
$\tau(S,\overline{A}).$ It is clear that if $S\simeq S^{\prime}$, then
$\tau(S)\simeq\tau(S^{\prime}$), \emph{i.e.,} that every structure in
$\tau(S)$\ is isomorphic to one in $\tau(S^{\prime}$), and vice-versa.

The formulas of the defining scheme of $\tau$ have free set variables
$Y_{1},...,Y_{m}$, intended to hold the $m$ parameters of $\overline{A}.$ The
precondition decides if these parameters are "correct" w.r.t. the intended construction.\ 

We say that $\tau$ is \emph{semantically deterministic} if any two structures
in $\tau(S)$ are isomorphic.\ This property can be guaranted by particular
constructions, but is not decidable from an arbitrary defining scheme.

\bigskip

The Backwards Translation Theorem holds, and the proof is the same as that of
Theorem 4.2, as it applies to structures $S$ expanded by $m$ additional unary
relations.\ We omit the details. This theorem yields \ the following one
(cf.\ \cite{CE}, Section 7.1.5 and Remark 7.15).\ 

\bigskip

\textbf{Theorem 4.4}\ : The composition of two MSO-transductions is an
MSO-transduction.\ The same holds for FO-transductions.\ 

\textbf{Proof sketch: }We consider the composition $\tau^{\prime}\circ\tau$ of
two FO-transductions $\tau^{\prime}$ and $\tau,$ respectively $q $- and
$p$-copying. The parameters of $\tau$ can be handled as unary relations of an
expansion of $S$.\ Hence, we can assume that $\tau$ has no parameter.\ 

Let $\tau$ transform $S$ into $T$\ and $\tau^{\prime}$ transform $T$ into
$U$.\ Hence $D_{T}\subseteq D_{S}\times\lbrack p]$ and $D_{U}\subseteq
D_{T}\times\lbrack q]\subseteq D_{S}\times\lbrack p]\times\lbrack q].$\ Hence
$\tau^{\prime}\circ\tau$ is $p.q$-copying. Hence $D_{U}$ is the union of sets
$D_{i,j}\times\{i\}\times\{j\}$ for $i\in\lbrack p]$, $j\in\lbrack q]$\ and
$D_{i,j}\subseteq D_{S}.$

Assume that $\tau^{\prime}$ has two parameters $X$ and $Y$ (to simplify the
proof).\ As subsets of $D_{T}\subseteq D_{S}\times\lbrack p]$ they are defined
from $2p$ subsets of $D_{S}$.\ Hence, $\tau^{\prime}\circ\tau$ uses the $2p$
parameters, $X_{i},Y_{i}$ for $i=1,...,p$.

An FO-definition scheme for $\tau^{\prime}\circ\tau$ consists of

1.\ a precondition $\chi^{\prime\prime}$ (depending on the set variables
$X_{i},Y_{i}$), of the form

$\chi\wedge\widehat{\chi^{\prime}}(X_{1},...,Y_{1},...$) where $\chi$ is the
precondition of $\tau$\ and $\widehat{\chi^{\prime}}$ is obtained from
$\chi^{\prime}$\ by backwards translation relative to $\tau$ (cf.\ Theorem 2.3(3)),

2. the\ $p.q$ domain formulas $\delta_{i,j}^{\prime\prime}(x)$ that define the
subsets $D_{i,j}$ of $D_{S}$; they are $\delta_{i}(x)\wedge\widehat
{\delta_{i,j}^{\prime}}(x,X_{1},...,Y_{1},...)$ where $\delta_{i}(x)$ defines
the subset $D_{i}$ of $D_{S}$\ forming the part $D_{i}\times\{i\}$ of $D_{T}$
and $\widehat{\delta_{i,j}^{\prime}}$ is obtained from $\delta_{i,j}^{\prime
}(x,X_{1},...,Y_{1},...)$ by backwards translation relative to $\tau,$ where
$\delta_{i,j}^{\prime}$ defines $D_{i,j}$ in $T$.\ 

The formulas defining the relations of $U$ are defined similarly.\ Backwards
translation does not introduce any set quantification. \hfill$\square$

\bigskip

The article \cite{BNOS}\ develops a thorough study of the quasi-order
$\sqsubseteq$ between graph classes such that $\mathcal{C}\sqsubseteq
\mathcal{D}$ if and only if $\mathcal{C}\subseteq\tau(\mathcal{D)}$ for some
FO-transduction $\tau$. However, their definition of FO-transductions is
weaker than ours as we now explain. It is given for undirected, simple
loop-free graphs, but this is not the main point.\ For our comparison, we use
below its straightforward extension to general relational structures. Here are
the definitions based on \cite{BNOS}.

\bigskip

\textbf{Definition 4.5 }\ : \emph{A weaker notion of linear expanding
FO-transduction}.

(a) The transduction $C_{k}$ transforms an $\mathcal{R}$-structure $S$\ into
the union of $k$ disjoint copies of it, keeping its relations in each
copy,\ together with a binary relation $Clone$ expressing that two elements
are copies of a same element of $D_{S}$. Hence $C_{k}(S)$ is an ($\mathcal{R}%
\cup\{Clone\}$)-structure and its domain is $D_{S}\times\lbrack k]$.

(b) For an $\mathcal{R}$-structure $S$ and an $m$-tuple $\mathcal{Y}$ of unary
relation names, intended to hold so-called \emph{colors} (without neighbouring
constraints), actually \emph{parameters} in our terminology,$\ \Gamma
_{\mathcal{Y}}(S)$ denotes the set of all ($\mathcal{R}\cup\mathcal{Y}%
$)-structures that expand $S$ with $m$ arbitrary subsets of its domain, that
give meanings to the relation names in $\mathcal{Y}$.

(c) An FO$^{-}$-transduction $\tau$ maps an $\mathcal{R}$-structure $S$ to the
set $I_{\Delta}(\Gamma_{\mathcal{Y}}(C_{k}(S))),$ where $\Delta$ is a fixed
FO-definition scheme without precondition (as in Definition 2.1), and
$\mathcal{Y}$ and $k$ are fixed as above. The signature of the output
structures is some $\mathcal{R}^{\prime}$ that need not contain $Clone$ and
$\mathcal{Y}$, which are used for applying $I_{\Delta}$.\ 

\bigskip

\textbf{Claim 4.6}\ : Every FO$^{-}$-transduction is (equivalent to) a linear
expanding FO-transduction without precondition.

\textbf{Proof : }The three basic transductions of Definition 4.5 are linearly
expanding FO-transductions.\ Their composition is of this form by Theorem 4.4.

An FO$^{-}$-transduction uses no precondition.\ This means that $I_{\Delta
}(\Gamma_{\mathcal{Y}}(C_{k}(S)))$ is defined for all choices of the
parameters in $\mathcal{Y}$.

We\ now compare $k$-copying FO- and FO$^{-}$-transductions.\ In an
FO-transduction, one choses parameters, say $A_{1},...,A_{n}$, \emph{before
copying} the input structure.\ For defining $\tau(S)$ by an FO$^{-}%
$-transduction, one choses $m$ parameters (colors) $B_{1},...,$ \ $B_{m}$
\ that are subsets of $D_{S}\times\lbrack k]$.\ For $i\in\lbrack m]$ and
$j\in\lbrack k]$, let $B_{i,j}$ be the set of $x\in D_{j}$ such that $(x,j)\in
B_{i}$\ where $D_{j}$ is the $j$-th copy of $D_{S}$\ in $C_{k}(S)$.\ These
sets are the $n:=m.k$ parameters of a corresponding FO-transduction.

The output structure of an FO$^{-}$-transduction involves the $Clone$ relation
that we can define by Definition 4.1.\ 

In a structure $C_{k}(S)$, there is no way to express that two elements are in
"a same copy" of $S$.\ One may argue that this does not matter as we consider
output structures up to isomorphism. (However, see Remark 5.5 below).

The notion of FO$^{-}$-transduction is weaker than that of an FO-transduction,
because of this last fact and of the absence of precondition.$\ $
\hfill$\square$

\bigskip

\textbf{Example 4.7\ }: \ Figure 1\ shows a graph $G$ consisting of the
edge\ $1-2$ and the clique $K_{3}$\ with vertices $3,4,5$.\ It shows
$C_{2}(G)$ to the right.\ Its vertices are $a,b,c,d,e$ for the first copy of
$1,2,3,4,5$\ and $a^{\prime},b^{\prime},c^{\prime},d^{\prime},e^{\prime}$, for
the second copy.\ However, $a,b,c^{\prime},d^{\prime},e^{\prime}$ can be
considered as forming the first copy and $a^{\prime},b^{\prime},c,d,e$ as the
second one.%

\begin{figure}
[ptb]
\begin{center}
\includegraphics[
height=2.4042in,
width=4.2497in
]%
{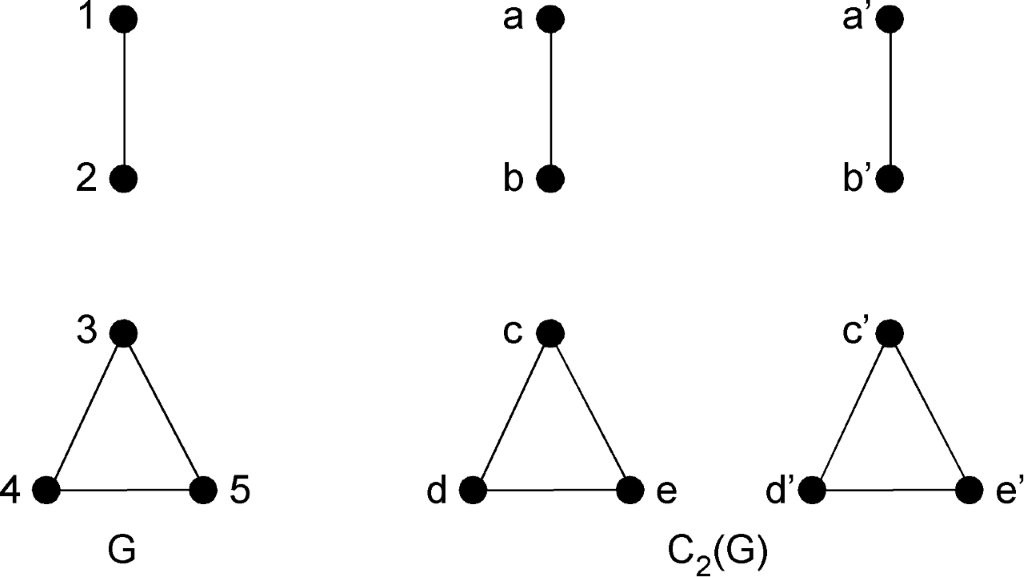}%
\caption{Example 4.7}%
\end{center}
\end{figure}

\bigskip

\textbf{Remark 4.8\ }: For an example, a non-copying FO$^{-}$--transduction
transforms the class interval graphs into that of all finite graphs, as proved
in \cite{BNOS}\ Example 10.1.\ This article shows many other cases whose
descriptions need a number of preliminary technical definitions.\ 

We think that its results\ remain valid with our notion of linearly expanding
FO-\ transduction in place of the FO$^{-}$ one, because the authors
investigate whether $\mathcal{C}\subseteq\tau(\mathcal{D)}$ (up to
isomorphism) and not $\mathcal{C}\simeq\tau(\mathcal{D)}$, so that the
preconditions of $\tau$ are not \ necessary in that case. A formal proof would
need a careful review the many technical proofs of \cite{BNOS}.

\section{Binary combinations of structures}

We consider disjoint union and notions of products based on different choices
of the defining formulas for the relations of the product structure.

\subsection{Disjoint union}

\textbf{Definitions 5.1 : }\emph{Disjoint union of relational structures.}

We recall from Definition 1.1 that if $S$ and $T$ are disjoint $\mathcal{R}%
$-structures\footnote{The extension to the case where $S$ and $T\ $have
different signatures and $S\oplus T$ is their union is straightforward.\ We
will use it below.
\par
{}}, their union, denoted by $S\oplus T$ (to insist on disjointness), has
domain $D_{S\oplus T}:=D_{S}\uplus D_{T}$ and relations $R_{S\oplus T}%
:=R_{S}\uplus R_{T}$ for all $R$ in $\mathcal{R}$. This operation is
commutative.\ The \emph{marked disjoint union} $S\oplus_{m}T$ produces an
$\mathcal{R}\cup\{In_{1},In_{2}\}$-structure where $In_{1}(x)$ (resp.\ $In_{2}%
(x))$ means that $x$ is in $D_{S}$ (resp.\ in $D_{T})$, hence, it is not
commutative\footnote{The mappings $S\longmapsto S\oplus S$ and $S\longmapsto
S\oplus_{m}S$ \ are 2-copying QF-transductions.}.

The following \emph{Splitting Theorem} is a key result.\ See \cite{Mak} for
its history, called there the \emph{Fefermann-Vaught Theorem}, and its
algorithmic applications.\ The algebraic aspect especially for MSO logic\ is
developped in \cite{CE}, Section 5.3. We will first state a special case,
extracted from Proposition 5.37 of this book.

\bigskip

We denote by $\sqcup$ the exclusive \emph{or} and by $\bigsqcup$ its extension
to several Booleans. Every Boolean expression can be converted into an
equivalent one of the form $\bigsqcup_{i\in I}\bigwedge\nolimits_{j\in K_{i}%
}b_{i,j}$ where $b_{i,j}$ is a Boolean variable or a negated one. The easiest
example is $a\vee b\Longleftrightarrow(a\wedge b)\sqcup(a\wedge\lnot
b)\sqcup(\lnot a\wedge b).$

\bigskip

We first give a particular case of the "full" \emph{Splitting Theorem} 5.4.

\bigskip

\textbf{Theorem 5.2 }: \ Let $\varphi$ be an FO sentence over $\mathcal{R}$.\ 

(1) One can construct an $s$-tuple of pairs of FO sentences $(\theta_{i}%
,\psi_{i})$ such that, for all disjoint $\mathcal{R}$-structures $S$ and $T$,
we have:

\begin{quote}
$[S\oplus T\models\varphi]$ $=\bigsqcup$ $_{1\leq i\leq s}\mathbb{[}%
S\models\theta_{i}]\wedge\lbrack T\models\psi_{i}],$
\end{quote}

where $[S\oplus T\models\varphi]$ is the Boolean equal to $\emph{true}$ if and
only if $S\oplus T$ $\models\varphi$ and similarly for the other sentences.

(2) The sentences $\theta_{i},\psi_{i}$ over $\mathcal{R}$\ have no larger
quantifier-height than $\varphi$.\ 

(3) The same holds with $\oplus_{m}$\ and a sentence $\varphi$ over
$\mathcal{R}\cup\{In_{1},In_{2}\}$.\ The sentences $\theta_{i},\psi_{i}$ are
over $\mathcal{R}$. \hfill$\square$

\bigskip

The formula $\varphi$ is decomposed (splitted) into sentences $\theta_{i}%
,\psi_{i}$ where $i\in\lbrack s].$ This decomposition does not depend on $S$
and $T$.\ A sentence $\theta_{i}$ may be false in some structure $S$ and
similarly for $\psi_{i}$ in $T$.

\bigskip

\textbf{Remark 5.3 : }(1) In an intermediate step of the proof, one builds a
$t$-tuple of pairs of sentences $(\theta_{i},\psi_{i})$ such that, for all
disjoint $\mathcal{R}$-structures $S$ and $T$, we have $[S\oplus
T\models\varphi]=\bigvee_{1\leq i\leq t}$ $\mathbb{[}S\models\theta_{i}%
]\wedge\lbrack T\models\psi_{i}].$ One can replace $\vee$\ by an exclusive
$or$ in the following way (Proposition 5.37 of \cite{CE}).\ For each nonempty
subset $A$\ of $[t]:=\{1,...,t\},$ one defines $\theta_{A}$ as the conjunction
of the $\theta_{i}$'s for $i$ in $A$\ and of the $\lnot\theta_{i}$'s for $i$
not in $A$.\ It follows that, if $A\neq A^{\prime}$, we cannot have
$S\models\theta_{A}\wedge\theta_{A^{\prime}}.\ $We define similarly $\psi_{B}$
for a nonempty subset $B$ of $[t]$.\ Then, the pairs $(\theta_{A},\psi_{B})$
such that $A\cap B\neq\emptyset$ (numbered appropriately) give the desired result.

We will use this technique below for Theorem 8.7.

(2) Theorem 5.2(1,2) extends to $S\oplus(T\oplus U)$\ isomorphic to ($S\oplus
T)\oplus U.$ For $S\oplus_{m}T\oplus_{m}U,$ we need $In_{3}$ in addition to
$In_{1}$ and $In_{2}$, for the elements from $U$.

(3) Consider now a binary operation $\boxplus$ on $\mathcal{R}$-structures
defined by $S\boxplus T:=\tau(S\oplus_{m}T)$ where $\tau$\ is a scalar
QF-transduction.\ (We have concrete examples of this form).\ Let $\varphi$\ be
a sentence over $\mathcal{R}$.\ Then:

\begin{center}
$S\boxplus T\models\varphi\ \Longleftrightarrow\ S\oplus_{m}T\models
\mu\ \Longleftrightarrow\ \bigsqcup_{1\leq i\leq s}\mathbb{[}S\models
\theta_{i}]\wedge\lbrack T\models\psi_{i}]=\mathit{true}$
\end{center}

where $\mu$\ is obtained from $\varphi$\ by Theorem 2.3(2)\ relative to
$\tau,$\ and $s,\theta_{i},\psi_{i}$ are obtained from $\mu$ by Theorem
5.2(3).\ Quantifier-heights do not increase, because $\tau$ is assumed
QF.\ \hfill$\square$

\bigskip

In the following generalization of Theorem 5.2, $\varphi$ has free variables
among $\{x_{1},...,x_{k},y_{1},...,y_{\ell}\}$, which we shortly denote by
$\overline{xy}$. We denote $(S\oplus T,\overline{a};\overline{b}%
)\models\varphi$ if $\varphi$ is satisfied in the usual sense, where
$\overline{a}$ $\in D_{S}^{k}$ and $\overline{b}$ $\in D_{S}^{\ell}$ give
values to $\overline{x}$ and $\overline{y}$. Formally, this is equivalent to
$(S\oplus_{m}T,\overline{a}\overline{b})\models\varphi\wedge In_{1}%
(\overline{x})\wedge In_{2}(\overline{y})$, where $In_{1}(\overline{x})$
abreviates $In_{1}(x_{1})\wedge...\wedge In_{1}(x_{k})$ and similarly for
$In_{2}(\overline{y})$.

\bigskip

\textbf{Theorem 5.4 (}\emph{Splitting Theorem for} $\oplus$\ \emph{and}
$\oplus_{m}$) : Let $\mathcal{R}$ be a relational signature.\ 

(1) Let $\varphi$ be an FO formula over $\mathcal{R}$\ with free variables in
the disjoint tuples $\overline{x}$\ and $\overline{y}$.\ One can construct a
finite set of FO\ formulas $(\theta_{i})_{i\in I}$ having their free variables
in $\overline{x}$, a finite set of FO\ formulas $(\psi_{i})_{i\in J}$ having
their free variables in $\overline{y}$ and a set $K\subseteq I\times J$ such
that, for all disjoint $\mathcal{R}$-structures $S $ and $T$, and all
$\overline{a}\in D_{S}^{k}$ and $\overline{b}$ $\in D_{S}^{\ell}$, we have

(a) $(S,\overline{a})\models\theta_{i}\wedge\theta_{j}$ implies $i=j$,

(b) $(T,\overline{b})\models\psi_{i}\wedge\psi_{j}$ implies $i=j$,

(c) $[S\oplus T,\overline{a};\overline{b}\models\varphi]=\bigvee
\mathbb{[}S,\overline{a}\models\theta_{i}]\wedge\lbrack T,\overline{b}%
\models\psi_{j}]$

where the disjunction extends to all pairs $(i,j)\in$ $K$. We also have :

(c') $[S\oplus T,\overline{a};\overline{b}\models\varphi]=\bigsqcup
\mathbb{[}S,\overline{a}\models\theta_{i}]\wedge\lbrack T,\overline{b}%
\models\psi_{j}]$

(2) The formulas $\theta_{i},\psi_{j}$ have no larger quantifier-height than
$\varphi.$

(3) The same holds for $\oplus_{m}$\ and $\varphi$ over $\mathcal{R}%
\cup\{In_{1},In_{2}\}$.\ The sentences $\theta_{i},\psi_{i}$ are over
$\mathcal{R}$.\ \hfill$\square$

\bigskip

Conditions (a) and (b) imply (c'),\emph{\ i.e.} that the disjunction in (c) is
exclusive. As in Theorem 5.2, the Boolean $\mathbb{[}S,\overline{a}%
\models\theta_{i}]$ expresses that we have $(S,\overline{a})\models\theta_{i}%
$, also written $S\models\theta_{i}(\overline{a}).$ We can have $S\simeq T,$
but $S$ and $T$ must be disjoint.\ No tuple in $R_{S\oplus T}$ can contain an
element of $S$ and one of $T$.

\bigskip

\textbf{Proof:} \ (1) The proof given in Lemma 5.36 and Proposition 5.37 of
\cite{CE} uses induction on the structure of $\varphi$. The formulas
$\theta_{i},\psi_{j}$ and the set $K$ depend only on $\varphi$ and the
partition of its allowed free variables into the tuples $\overline{x}$ and
$\overline{y}$. (A variable in $\overline{x}$ may actually not occur in
$\varphi$).

We review the case of $\varphi$ of the form $\exists u.\varphi^{\prime}.$ By
induction, we have two splittings for $\varphi^{\prime}$: the one intended for
finding $u$ in $D_{S}$ so that, for all $\overline{a}\in D_{S}^{k}$ , $c\in
D_{S}$ and $\overline{b}$ $\in D_{S}^{\ell}$ we have

\begin{quote}
$[S\oplus T,\overline{a}c;\overline{b}\models\varphi^{\prime}]=\bigsqcup
_{(i,j)\in K}\mathbb{[}S,\overline{a}c\models\theta_{i}]\wedge\lbrack
T,\overline{b}\models\psi_{j}]\ \ \hfill(1)$
\end{quote}

and the one intended for finding in $D_{T}$ a value of $u$ so that, for all
$\overline{a},$ $\overline{b}$ as above and $c\in D_{T}$, we have

\begin{quote}
$[S\oplus T,\overline{a};\overline{b}c\models\varphi^{\prime}]=\bigsqcup
_{(i,j)\in K^{\prime}}\mathbb{[}S,\overline{a}\models\theta_{i}^{\prime
}]\wedge\lbrack T,\overline{b}c\models\psi_{j}^{\prime}].\ \hfill(2)$
\end{quote}

Then we get :

\begin{quote}
$[S\oplus T,\overline{a};\overline{b}\models\exists u.\varphi^{\prime
}]=\bigsqcup_{(i,j)\in K}\mathbb{[}S,\overline{a}\models\exists u.\theta
_{i}]\wedge\lbrack T,\overline{b}\models\psi_{j}]\vee$

$\qquad\qquad\qquad\qquad\qquad\bigsqcup_{(i,j)\in K^{\prime}}\mathbb{[}%
S,\overline{a}\models\theta_{i}^{\prime}]\wedge\lbrack T,\overline{b}%
\models\exists u.\psi_{j}^{\prime}].$ \ \ \ \ (3)
\end{quote}

The expression (3) contains $\vee$. We can obtain an expression with exclusive
$or$ as said in Remark 3.3, by noting that $\lnot\mathbb{[}S,\overline
{a}\models\exists u.\theta_{i}]\Longleftrightarrow$ $\mathbb{[}S,\overline
{a}\models\lnot\exists u.\theta_{i}]$, that $\lnot\lbrack T,\overline
{b}\models\psi_{i}]\Longleftrightarrow$ $\mathbb{[}T,\overline{b}\models
\lnot\psi_{i}]$ and that $\mathbb{[}S,\overline{a}\models\exists u.\theta
_{i}]\wedge\mathbb{[}S,\overline{a}\models\theta_{i}^{\prime}%
]\Longleftrightarrow$ $\mathbb{[}S,\overline{a}\models(\exists u.\theta
_{i})\wedge\theta_{i}^{\prime}]$ etc.

It follows that the expression (3) can be rewritten into an exclusive
disjunction of conjunctions of the form $\mathbb{[}S,\overline{a}\models
\alpha]\wedge\mathbb{[}T,\overline{b}\models\beta].\ \ \ \ $

(2) and (3$),$ by inspecting the proofs. \hfill$\square$

\bigskip

\textbf{Remark 5.5}\ : \emph{On the power of marked union}.

Let $\mathcal{C}$ be the class of undirected graphs $G=(V,edg,A,B)$ such that
$A$ and $B$ are subsets of $V$ (handled as unary relations).\ Let
$\tau:\mathcal{C}\rightarrow\mathcal{C}$ be the QF-transduction that adds to a
graph $G$ edges between all vertices of $A$ and all vertices of $B$. There
exists a QF-transduction $\tau^{\prime}$ that maps $G\oplus_{m}H$\ to
$\tau(G)\oplus\tau(H)$ for any two disjoint $G,H$ in $\mathcal{C}$. However,
there is no FO- (and even no MSO-) transduction that maps $G\oplus H $\ to
$\tau(G)\oplus\tau(H)$. The reason is that in $G\oplus H$\ one cannot
distinguish the vertices in $A$ from those in $B$.

\bigskip

\textbf{Definition 5.6} (\cite{CE}, Definition 2.128): \emph{Parallel
composition of relational structures with sources.}

Let $S$ and $T$ be disjoint $\mathcal{R}$-structures, where $\mathcal{R}$
contains unary relations $Eq_{a}$, ..., $Eq_{d}$ intended to denote constants
$a_{S},...,d_{S}$\ in $D_{S}$\ and similarly in $T$, cf. Definition 1.1, that
we call \emph{sources, }designated by their\emph{\ names, }$a,...,d$%
\emph{.\ }These relations must be singleton or empty. We assume that no two
constants among $a,...,d$ name a same element of $S$ or of $T$. We do not
assume that $S$ and $T$ have exactly the same named sources.\ Then $S//T$\ is
the $\mathcal{R}$-structure $S\oplus T$ where the sources with a same name in
$S$ and $T$ are fused.

For an example $S//T=(a\rightarrow b\rightarrow\ast\rightarrow c)$ if
$S=(a\rightarrow b)$ and $T=(b\rightarrow\ast\rightarrow c).$ The operations
// and $\bullet$ that define series-parallel graphs form a very classical
example, see \cite{CE}, Section 1.1.3 for their descriptions in a logical setting.

\bigskip

\textbf{Proposition} \textbf{5.7}\ : (1) For $\mathcal{R},S,T$ and source
names $a,...,d$ as above, we have $S//T=\tau(S\oplus_{m}T)$ for some fixed
scalar FO-transduction.

(2) We have a splitting theorem for $//$, and fixed sets of existing sources
in the two argument structures, however the quantifier-heights increase in the decomposition.

\textbf{Proof} : (1) We let $\tau$ delete from $S\oplus_{m}T$ the elements $x$
such that $In_{2}(x)\wedge Eq_{a}(x)$ holds where $a_{S}$ is a source of $S$.
Hence, $\delta(x)$ is a QF formula depending on the exact sets of sources of
$S$ and $T$.

For simplicity, we consider that $\mathcal{R}$\ consists only of the binary
relation $edg$ and the relations $Eq_{a},...,Eq_{d}$.\ Then $edg(x,y)\ $is
defined in $S//T$\ by the following FO formula\ :

$\delta(x)\wedge\delta(y)\wedge$

{\large (}$edg(x,y)\vee\bigvee\nolimits_{a}[\{\exists z.(In_{2}(z)\wedge
Eq_{a}(z)\wedge edg(z,y))\wedge Eq_{a}(x)\}$

$\qquad\qquad\qquad\qquad\qquad\vee\{\exists z.(In_{2}(z)\wedge Eq_{a}%
(z)\wedge edg(x,z))\wedge Eq_{a}(y)\}]${\large )},

where the disjunction extends to all nullary symbols $a,...,d$. \ Hence $\tau$
is not a QF-transduction.

(2) Immediate consequence of Theorem 4.2(3) and Theorem 5.4(3). \hfill
$\square$

\bigskip

However, by using nullary symbols for sources rather than unary relations
$Eq_{a}$,..., $Eq_{d}$,\ we could define $\theta$\ as a QF-transduction with constants.\ 

In particular, $In_{2}(x)\wedge Eq_{a}(x)$ would be replaced by $In_{2}%
(x)\wedge x=a,$ and

$\exists z.(In_{2}(z)\wedge Eq_{a}(z)\wedge edg(z,y))\wedge Eq_{a}(x)$ by
$edg(a,y)\wedge x=a.$

In this case, we have an analog of Proposition 5.7(2) without increase of
quantifier-heights.\ This is important for proving the recognizability of the
class of finite models of a first-order sentence, see Section 7\ below.

\bigskip

\subsection{Products of disjoint structures}

\bigskip

A \emph{product} of two relational structures $S$ and $T$ has domain
$D_{S}\times D_{T}$ and relations defined in different possible ways.\ We
first give definitions and examples for graphs.

\bigskip

\textbf{Definitions 5.8}\ : \emph{Products of simple directed graphs}

We consider directed graphs $G$ described by structures $(V_{G},edg_{G})$
where $edg_{G}$ is a binary relation. Here are some cases among many others,
described with the notation of \cite{Wiki}.

\bigskip

\emph{Cartesian product}:

$G\square H\ =(V_{G}\times V_{H},edg_{G\square H\ })$ where

$edg_{G\square H\ }((x,y),(x^{\prime},y^{\prime})):\Longleftrightarrow
(x=x^{\prime}\wedge edg_{H}(y,y^{\prime}))\vee(edg_{G}(x,x^{\prime})\wedge
y=y^{\prime})).$

\bigskip

\emph{Kronecker (or categorical) product}

$G\times H$ where $\ edg_{G\times H\ }((x,y),(x^{\prime},y^{\prime
})):\Longleftrightarrow edg_{G}(x,x^{\prime})\wedge edg_{H}(y,y^{\prime}).$

\bigskip

\emph{Lexicographical product}

$G[H]$ where $edg_{G[H]\ }((x,y),(x^{\prime},y^{\prime})):\Longleftrightarrow
edg_{G}(x,x^{\prime})\vee(x=x^{\prime}\wedge edg_{H}(y,y^{\prime})).$

\bigskip

In all cases, the existence of an edge between $(x,y)$ and $(x^{\prime
},y^{\prime})$ is defined by a Boolean combination of $x=x^{\prime
},y=y^{\prime},edg_{G}(x,x^{\prime})$ and $edg_{H}(y,y^{\prime}).$ We will
consider products of relational structures that subsume all these cases.\ 

\bigskip

\textbf{Examples 5.9 \ }: \emph{Some products of directed graphs.}

We consider structures representing directed graphs. Here $P_{n}$ denotes the
directed path $1\rightarrow2\rightarrow...\rightarrow n$. The product $P_{n}$
$\square P_{n}$ is the $n\times n$ directed square grid $G_{n\times n} $. The
graphs $P_{n}$ $\times P_{n}$ have tree-width 2, whereas $P_{n}$ $\square
P_{n}$ and $P_{n}[P_{n}]$\ have unbouded tree-width.\ 

Figure 2 shows\ $P_{4}$ $\square P_{4}$ and $P_{4}$ $\times P_{4}$. The
product $P_{4}$ $\times P_{4}$ consists of 5 directed paths and 2 isolated
vertices. Figure 3 shows $P_{4}[P_{4}]$ that consists of 4 directed paths
connected by 3 directed $K_{4,4}$ graphs, summarized by $\overrightarrow
{\otimes}.$

Note that the graphs $P_{4}\square P_{4},P_{4}$ $\times P_{4}$ are subgraphs
of $P_{4}[P_{4}]$. \hfill$\square$%
\begin{figure}
[ptb]
\begin{center}
\includegraphics[
height=2.271in,
width=5.9577in
]%
{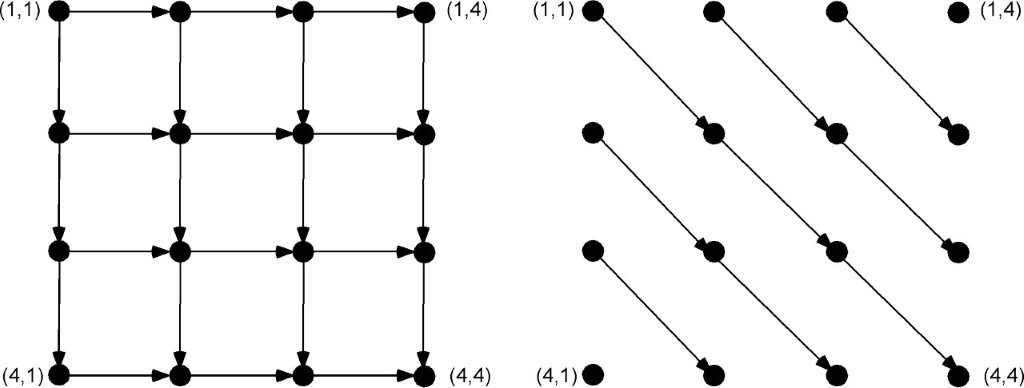}%
\caption{$P_{4}\square P_{4}$ and $P_{4}\times P_{4}$}%
\end{center}
\end{figure}
%

\begin{figure}
[ptb]
\begin{center}
\includegraphics[
height=2.3514in,
width=2.962in
]%
{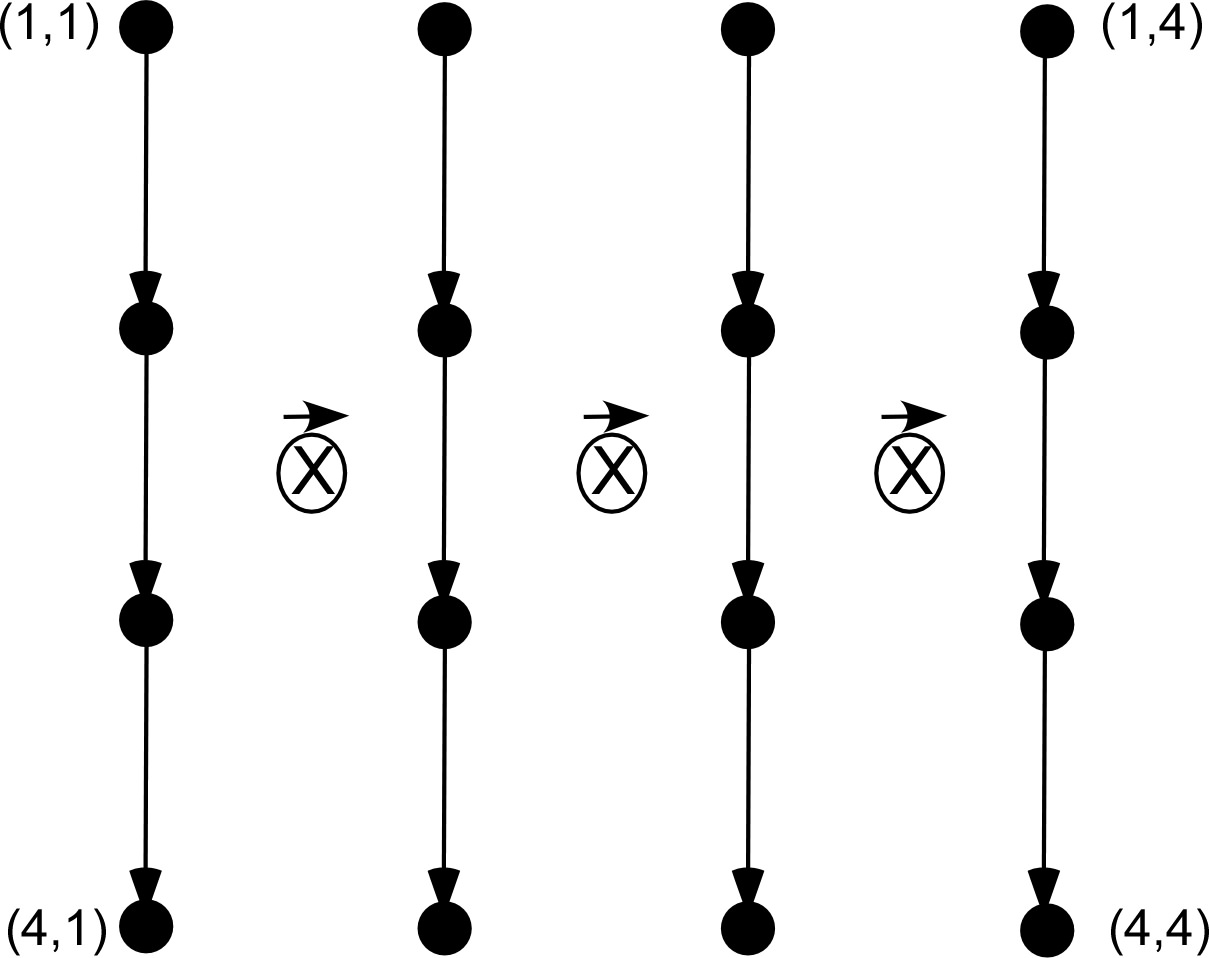}%
\caption{$P_{4}[P_{4}]$ with directed $K_{4,4}$ represented by
$\overrightarrow{\otimes}$}%
\end{center}
\end{figure}

\bigskip

\textbf{Definition 5.10 }: \emph{Products of relational structures}

A \emph{product} of two disjoint relational structures $S$ and $T$, possibly
over different signatures, respectively $\mathcal{R}$ and $\mathcal{R}%
^{\prime}$, has domain $D_{S}\times D_{T}$ and relations defined in different
possible ways. In view of Lemma 5.12\ below, it is convenient to assume that
$D_{S}$ and $D_{T}$\ are disjoint. If they are not, as in Examples 5.9, then
one replaces one by an isomorphic copy disjoint from the other.

Such a product over a signature $\mathcal{R}^{\prime\prime}$\ (possibly
neither $\mathcal{R}$ nor $\mathcal{R}^{\prime}$ ) is denoted by
$S\times_{\sigma}T$ where $\sigma$ \ is a tuple of defining formulas (somewhat
similar to the definition schemes of Definitions 2.1, 4.1 and 4.3).

It is a \emph{QF-product} if its relations are defined by QF\ formulas (as in
Examples 5.9).\ It is an \emph{FO-product} if its relations are defined by FO formulas.

An $n$-ary relation $U(z_{1},...,z_{n})$ of $S\times_{\sigma}T$, where
$z_{i}:=(x_{i},y_{i})$, is defined by an FO- (or QF-) formula $\sigma
_{U}(x_{1},...,x_{n},y_{1},...,y_{n})$ whose atomic formulas are of the forms
$x_{i}=x_{j}$, $y_{i}=y_{j}$, $R(u,...,w)$, $R^{\prime}(u^{\prime
},...,w^{\prime})$ where $R$ $\in$ $\mathcal{R}$, $R^{\prime}\in
\mathcal{R}^{\prime}$, $u,...,w\in\{x_{1},...,x_{n}\}$ and $u^{\prime
},...,w^{\prime}\in\{y_{1},...,y_{n}\}$. The variables $x_{i}$ will only take
values in $D_{S}$ and the $y_{j}$'s in $D_{T}$.\ As in Definition 5.1, the
unary relations $In_{1}(u)$ and $In_{2}(u)$ can impose that $u$ takes its
values in $S$ and, respectively, in $T$.\ Formally,

\begin{quote}
$U_{S\times_{\sigma}T}(c_{1},...,c_{n}):\Longleftrightarrow$

$\qquad(S\oplus T,(a_{1},...,a_{n},b_{1},...,b_{n}))\models\sigma
_{U}(\overline{x},\overline{y})\wedge In_{1}(\overline{x})\wedge
In_{2}(\overline{y})$

\qquad where $c_{i}=(a_{i},b_{i})$, $\overline{x}=(x_{1},...,x_{n})$ and
$\overline{y}=(y_{1},...,y_{n}).$ \hfill$\square$
\end{quote}

\bigskip

For each of these products, we have a splitting theorem, cf. \cite{CE,Mak},
similar to Theorems 5.2 and 5.4, see below Theorem 5.13

\bigskip

\textbf{Example\ 5.11 }: \emph{A toroidal grid.}

For an example, we consider\ the toroidal grid $T_{3,4}$ defined as the
rectangular grid $P_{3}\square P_{4}$ augmented with the arcs
$(4,i)\rightarrow(1,i)$ for $i:=1,2,3$ and $(i,3)\rightarrow(i,1)$ for
$i:=1,...,4.\ $See Figure 4.%

\begin{figure}
[ptb]
\begin{center}
\includegraphics[
height=3.0562in,
width=3.2655in
]%
{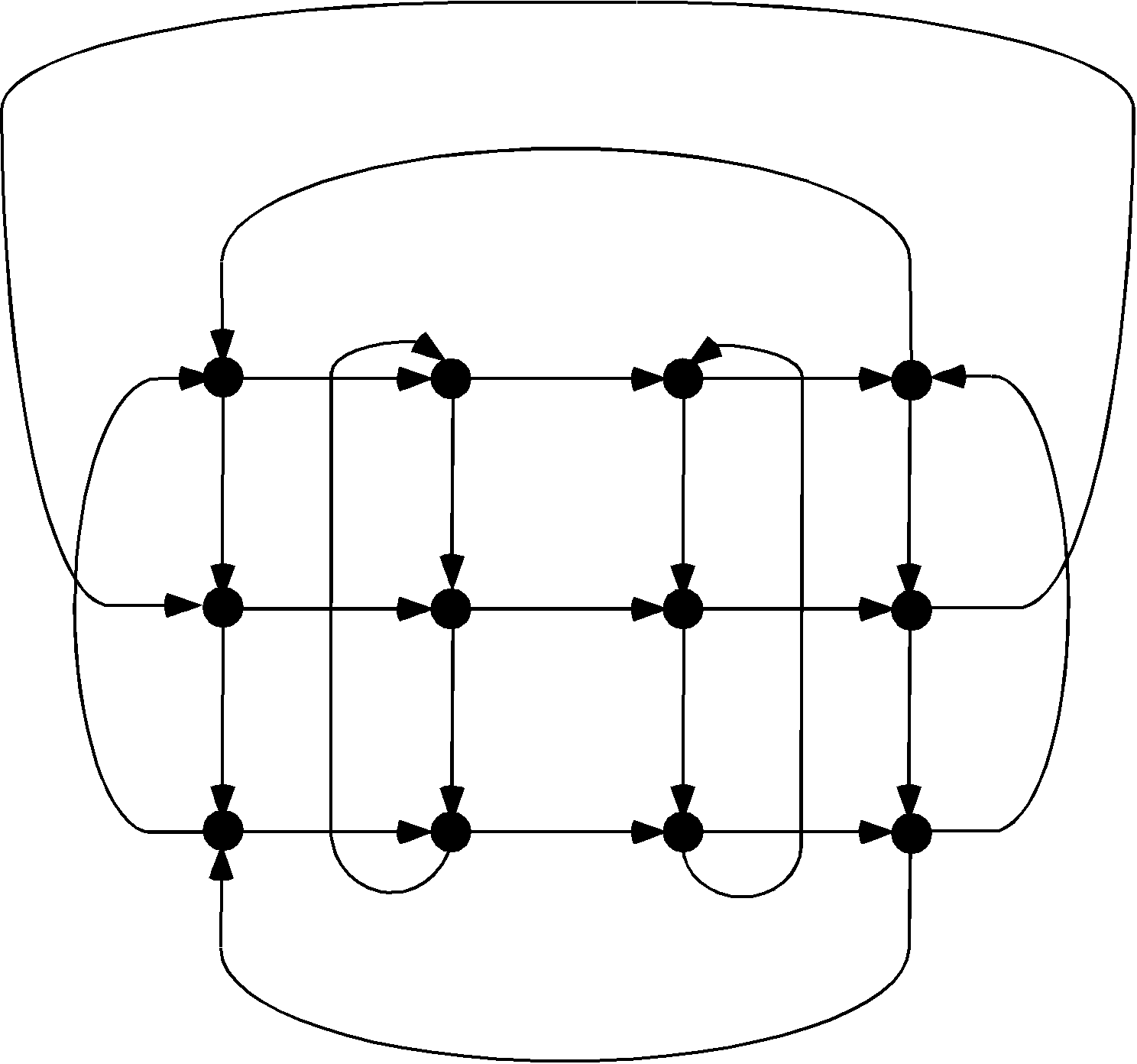}%
\caption{The toroidal grid $T_{3,4.}$}%
\end{center}
\end{figure}

\bigskip

The definition of the toroidal grid $T_{n,m}=T$\ based on $P_{n}\square P_{m}
$ (with $n,m\geq2$) is given below where a path is represented by the
structure $(N,\rightarrow,First,Last)$ where $N$ is the set of positions,
$\rightarrow$\ is the successor relation, $First$ and $Last$ are unary
relations defining respectively the first and the last position. In $T$, we have:

\begin{quote}
$edg_{T}((x,y),(x^{\prime},y^{\prime}))$ if and only if

$\{In_{1}(x)\wedge In_{1}(x^{\prime})\wedge In_{2}(y)\wedge In_{2}(y^{\prime
})\}\wedge$

$\{[x=x^{\prime}\wedge\lbrack y\rightarrow y^{\prime}\vee(Last(y)\wedge
First(y^{\prime}))]]\vee$

\qquad$\lbrack y=y^{\prime}\wedge\lbrack x\rightarrow x^{\prime}%
\vee(Last(x)\wedge First(x^{\prime}))]]\}.$
\end{quote}

If the composing paths are defined by $(N,\rightarrow)$ then the defining
formulas are FO\ and not QF, because $First\ $ and $Last\ $are not
QF-definable.\ \hfill$\square$

\bigskip

Let $\mathcal{R}$ and $\mathcal{R}^{\prime}$ be signatures, let $\times
_{\sigma}$ be an FO-product between $\mathcal{R}$-structures and
$\mathcal{R}^{\prime}$-structures producing $\mathcal{R}^{\prime\prime}$
structures. Let $\varphi$ be an FO formula over $\mathcal{R}^{\prime\prime}$
with free variables in $\overline{z}:=(z_{1},...,z_{k})$.\ Each $z_{i}$ is a
pair $(x_{i},y_{i})$ taking its values in $D_{S}\times D_{T}$.\ If
$\overline{c}$ is a $k$-tuple in $D_{S}\times D_{T}$, then $pr_{1}%
(\overline{c})$ denote its projection onto a $k$-tuple in $D_{S}$ and
$pr_{2}(\overline{c})$ similarly in $D_{T}$.\ Hence, if $\overline{c}%
=((a_{1},b_{1}),...,(a_{k},b_{k}))$, we have $pr_{1}(\overline{c}%
)=(a_{1},...,a_{k})=\overline{a}$\ \ and $pr_{2}(\overline{c})=(b_{1}%
,...,b_{k})=\overline{b}$.

\bigskip

\textbf{Lemma 5.12 }: Let $\mathcal{R}$, $\mathcal{R}^{\prime}$,
$\times_{\sigma}$, $\mathcal{R}^{\prime\prime}$ and $\varphi$ be as
above.\ Hence $\varphi$ has free variables in $\overline{z}=((x_{1}%
,y_{1}),...,(x_{k},y_{k}))$, hence actually in $\{x_{1},y_{1},...,x_{k}%
,y_{k}\}$.

(1) One can construct an FO\ formula $\widehat{\varphi}(x_{1},...,x_{k}%
,y_{1},...,y_{k})$ such that for every $\mathcal{R}$-structure $S$,
$\mathcal{R}^{\prime}$-structure $T$ and $k$-tuple $\overline{c}$ in
$(D_{S}\times D_{T})^{k}$, we have

\begin{quote}
$(S\times_{\sigma}T,\overline{c}\models\varphi)\Longleftrightarrow(S\oplus
T,pr_{1}(\overline{c});pr_{2}(\overline{c}))\models\widehat{\varphi}.$
\end{quote}

(2) If $\times_{\sigma}$is a QF-product, then the quantifier-height of
$\widehat{\varphi}$\ is at most twice that of $\varphi.$

\textbf{Proof} : (1) As usual by induction on the structure of $\varphi.$ Here
are the main cases.

If $\varphi$ is $\overline{z}=\overline{z}^{\prime}$, then $\widehat{\varphi}$
is $x_{1}=x_{1}^{\prime}\wedge...\wedge x_{k}=x_{k}^{\prime}$ $\wedge$
$y_{1}=y_{1}^{\prime}\wedge...\wedge y_{k}=y_{k}^{\prime}$.

If $\varphi$ is $U(\overline{z}_{1},...,\overline{z}_{k}),$ then
$\widehat{\varphi}$ is $\sigma_{U}(x_{1},...,x_{k},y_{1},...,y_{k}).$

If $\varphi$ is $\exists z_{k+1}.\varphi_{1},$ then $\widehat{\varphi}$ is
$\exists x_{k+1},y_{k+1}.\widehat{\varphi_{1}}.$

(2) Clear from the inductive construction.\ \hfill$\square$

\bigskip

The next theorem follows from Theorem 5.4.

\bigskip

\textbf{Theorem 5.13 }: Let $\mathcal{R}$, $\mathcal{R}^{\prime}$,
$\times_{\sigma}$ and $\mathcal{R}^{\prime\prime}$ and $\varphi$\ be as in
Lemma 5.12, hence, $\times_{\sigma}$ is an FO-product.\ 

(1) One can construct an $s$-tuple of pairs of formulas $(\theta_{i},\psi
_{i})$ such that $\theta_{i}$ (over $\mathcal{R)}$ has free variables in
$\overline{x}=(x_{1},...,x_{k})$, and $\psi_{i}$ (over $\mathcal{R}^{\prime})$
has free variables in $\overline{y}$, and for every $\mathcal{R}$-structure
$S$, $\mathcal{R}^{\prime}$-structure $T$ and $k$-tuple $\overline{c}$ in
$(D_{S}\times D_{T})^{k}$, we have

\begin{quote}
$[S\times_{\sigma}T,\overline{c}\models\varphi]=\bigsqcup_{1\leq i\leq
s}\mathbb{[}S,pr_{1}(\overline{c})\models\theta_{i}]\wedge\lbrack
T,pr_{2}(\overline{c})\models\psi_{i}].$
\end{quote}

(2) If $\times_{\sigma}$ is a QF-product, the formulas $\theta_{i},\psi_{i}$
have no\ larger quantifier-height than $\varphi$.

\bigskip

\textbf{Proof }: By Lemma 5.12(1),\ we have

\begin{quote}
$(S\times_{\sigma}T,\overline{c})\models\varphi\Longleftrightarrow(S\oplus
T,pr_{1}(\overline{c});pr_{2}(\overline{c}))\models\widehat{\varphi}.$
\end{quote}

Then, we get the result by decomposing $\widehat{\varphi}$ with Theorem 5.4.

Let us examine the case of $\varphi$ of the form $\exists z_{k+1}.\varphi_{1}%
$.\ Then $\widehat{\varphi}$ is $\exists x_{k+1},y_{k+1}.\widehat{\varphi_{1}%
}.$ We consider $(S\times_{\sigma}T,\overline{c})\models\varphi.$ Let
$pr_{1}(\overline{c}):=\overline{a}$ and $pr_{2}(\overline{c}):=\overline{b}.$
We have

\begin{quote}
$(S\times_{\sigma}T,\overline{c})\models\exists z_{k+1}.\varphi_{1}$ if and
only if \ 

there exist $d\in D_{S}$\ and $f\in D_{T}$ such that $(S\oplus T,\overline
{a}d;\overline{b}f)\models\widehat{\varphi_{1}}$.
\end{quote}

By Theorem 5.4\ this is equivalent to :

\begin{quote}
$\bigsqcup\mathbb{[}S,\overline{a}d\models\theta_{i}]\wedge\lbrack
T,\overline{b}f\models\psi_{j}]$
\end{quote}

where the disjunction extends to all pairs $(i,j)\ $in some set $K,$ and the
formulas $\theta_{i}$ and $\psi_{j}$ have no larger quantifier-height than
$\varphi.$ Hence $(S\times_{\sigma}T,\overline{c})\models\exists
z_{k+1}.\varphi_{1}$ is equivalent to

\begin{quote}
$\bigsqcup\mathbb{[}S,\overline{a}\models\exists x_{k+1}\theta_{i}%
]\wedge\lbrack T,\overline{b}\models\exists y_{k+1}\psi_{j}].$
\end{quote}

where $K,\theta_{i}$ and $\psi_{j}$ are as above.

(2) The quantifier-height\ does not increase (as in Lemma 5.12(2)) because the
decomposition of $\exists x_{k+1},y_{k+1}.\widehat{\varphi_{1}}$ based on
Theorem 5.4\ distributes\ the increments on separate formulas. \hfill$\square$

\bigskip

\textbf{Remark 5.14\ }: Because of Lemma 5.12, one might think that
$S\times_{\sigma}T$ is obtained by an FO\ (or an MSO) transduction from
$S\oplus_{m}T$.\ This is not the case because the cardinality of
$S\times_{\sigma}T $ is not $O(\left\vert D_{S}\uplus D_{T}\right\vert )$.\ 

Note also that $P_{n}$ $\square P_{n}$ of Example 5.9 is the grid $G_{n\times
n}$, that $P_{n}$ $\oplus P_{n}$ has tree-width 2 and the grids $G_{n\times
n}$'s have unbounded tree-width, whereas MSO\ transductions preserve bounded
tree-width. See \cite{CE}, Chapter 7 for details.

\bigskip

\textbf{Remark 5.15}\ : \emph{A linear expanding FO-transduction can be seen
as a product with a fixed structure.}

For each finite positive integer $k$, we let \textbf{[}$k$\textbf{]} be the
structure $([k],Eq_{1},...,Eq_{k})$ where $Eq_{1},...,Eq_{k}$ are unary
relations such that $Eq_{i}(x):\Longleftrightarrow x=i$. Hence, we make each
$i=1,...,k$ into a constant of the structure\ \textbf{[}$k$\textbf{]} while
keeping a relational signature (cf.\ Definition 1.1).

We now consider FO and QF\ products of the form $T:=S\times_{\sigma}%
$\textbf{[}$k$\textbf{]} according to Definition 5.10.\ Hence $D_{T}%
=D_{S}\times\lbrack k].$ In each relation $U_{T}(c_{1},...,c_{n}),$ we have
$c_{i}=(a_{i},j_{i})$ where $j_{i}\in\lbrack k]$. Hence $U_{T}$ is composed of
$k$-ary relations on $D_{S}$ such that:

\begin{quote}
$\overline{U}_{j_{1},...,j_{k}}(a_{1},...,a_{k})$ $\Longleftrightarrow$
$((a_{1},j_{1}),...,(a_{k},j_{k}))\in U_{T}.$
\end{quote}

\bigskip

\emph{Claim\ :} Each relation $\overline{U}_{j_{1},...,j_{k}}$\ is
QF-definable (resp. FO-definable) in $S$ if $\times_{\sigma}$\ is a QF-product
(resp.\ an FO-product).\ Conversely, QF-definable (resp. FO-definable)
relations $R_{j_{1},...,j_{k}}$\ \ can be combined into a quantifier-free
(resp. a first-order) formula $\theta_{U}$\ giving\ $\overline{U}%
_{j_{1},...,j_{k}}\ =R_{j_{1},...,j_{k}}\ $.\ The corresponding
transformations of formulas are effective.

Hence, linear expanding transductions, can be seen as products of structures
without particular structures.\ 

\section{Vectorial transductions}

We consider transductions $S$ $\mapsto T$\ such that $D_{T}\subseteq D_{S}%
^{k}$ for some fixed integer $k\geq2$.

\bigskip

\textbf{Definition 6.1 }: \emph{Vectorial FO- and QF-transductions.}

Let $\mathcal{R}$ and $\mathcal{R}^{\prime}$ be relational signatures and
$k\geq2$. A \emph{vectorial FO-transduction of dimension k }maps $S$ $\mapsto
T$\ such that $D_{T}\subseteq D_{S}^{k}$ and$\ D_{T}=\{(x_{1},...,x_{k})\in
D_{S}^{k}\mid S\models\delta(x_{1},...,x_{k})\}$ for some fixed FO\ formula
$\delta$ over $\mathcal{R}$.\ We let $\overline{z}_{i}=(z_{i,1},...,z_{i,k}).$
Then for an $n$-ary relation $U$ in $\mathcal{R}^{\prime}$, we define
$U_{T}(\overline{z}_{1},...,\overline{z}_{n})$ by $S\models\theta_{U}%
(z_{1,1},...,z_{1,k},...,z_{n,1},...,z_{n,k})$ where the $\theta_{U}$'s are
fixed FO\ formulas over $\mathcal{R}$.

A \emph{definition scheme} is here a tuple $\Delta:=(\delta,\theta
_{U},...,\theta_{U^{\prime}}).$ If all formulas are QF, we have a
\emph{QF-transduction}.

An auxiliary linear order may help, as we will see in Examples 6.2(b,c).\ We
get the notion of an \emph{order-invariant vectorial FO-transduction}.\ We can
also allow set parameters, as in Definitions 2.1(b) and 4.3, and then use FO
expressible preconditions.\ 

\bigskip

\textbf{Examples 6.2 : }We give examples of vectorial QF-transductions of
dimension 2, relative to graphs.

(a) For a simple loop-free directed graph $G$ defined as $(V,edg)$ (it has no
two parallel edges), we let $\left\lceil G\right\rceil $ be its
\emph{incidence structure} whose domain is $V\uplus E,$\ where $E$ is the set
of arcs.\ That is $E$ is the set of pairs $(x,y)$ such that $edg(x,y)$ holds.
The \emph{incidence relation} $inc(u,v)$ is defined as follows:

\begin{quote}
either $v$ is the arc $(u,w)$ or $u$ is the arc $(w,v)$, for some vertex $w.$
\end{quote}

We build $\left\lceil G\right\rceil $ with domain included in $V\times V$\ by
using $(x,x)$ to represent a vertex $x$.\ (This is possible since $G$ has no
loop). The domain formula is defined by : $\delta(x,y):\Longleftrightarrow
x=y\vee edg(x,y),$\ where the condition $x=y$ expresses that each pair $(x,x)$
is allowed in order to represent a vertex. Then $inc((x,y),(x^{\prime
},y^{\prime}))$ is defined by

\begin{quote}
$(x=y=x^{\prime}\wedge edg(x^{\prime},y^{\prime}))\vee(edg(x,y)\wedge
x^{\prime}=y=y^{\prime})$.
\end{quote}

(b) If $G$ is simple loop-free and undirected, then $edg(x,y)$
$\Longleftrightarrow$ $edg(y,x)$.\ The edges in $G$\ are defined in
$\left\lceil G\right\rceil $ as unordered pairs of adjacent vertices.\ In this
case, $inc(u,v)$ holds if and only if $u$ is an edge and $v$ is one of its
vertices. (cf. \cite{CE},\ Definition 5.17).

For building $E$ as a subset of $V\times V$, we must choose between
$(x,y)$\ and $(y,x)$ when $x$ and $y$ are adjacent \ for defining the
corresponding edge. An auxiliary linear order is\ here useful: one chooses
$(x,y)$ if $x<y$.\ For different linear orders, any two produced structures
$\left\lceil G\right\rceil $ are isomorphic.\ 

(c) The line graph $Line(G)$ for $G$ as in (b) has vertex set $E$, where $e,f
$ are adjacent if these two edges share a vertex. We can construct $Line(G)$
as follows:

\begin{quote}
$(G,\leq)\longmapsto\left\lceil G\right\rceil \mapsto Line(G)$
\end{quote}

where the second transduction deletes the pairs $(x,x)$ and defines (we have
$x<y$ and $x^{\prime}<y^{\prime}),$ then

\begin{center}
$edg_{Line(G)}((x,y),(x^{\prime},y^{\prime})):\Longleftrightarrow
(x,y)\neq(x^{\prime},y^{\prime})\wedge(\{x,y\}\cap\{x^{\prime},y^{\prime
}\}\neq\varnothing).$
\end{center}

(d) The product $P_{n}\square P_{n}$ where $P_{n}$ is the path
$1\longrightarrow2\longrightarrow...\longrightarrow n$ defines a\ square
directed grid $G_{n\times n}$ (cf.\ Figure 2).\ By adding the diagonal path
$(1,1)\longrightarrow(2,2)\longrightarrow...\longrightarrow(n,n)$ we get
$D_{n,n}$.\ For an arbitrary directed path $P=(V,edg_{P})$, the associated
\emph{augmented grid} $D$\ is defined from $V\times V$ as follows:

\begin{quote}
$edg_{D}((x,y),(x,y))$ if and only if

$(x=x^{\prime}\wedge edg_{P}(y,y^{\prime}))\vee(edg_{P}(x,x^{\prime})\wedge
y=y^{\prime})\vee$

$\qquad\qquad\qquad\qquad(edg_{P}(x,x^{\prime})\wedge x=y\wedge x^{\prime
}=y^{\prime})$.
\end{quote}

The first two alternatives of this last QF definition specify $edg_{P_{n}%
\square P_{n}}$; the third one defines the diagonal path. The transduction
$P\longmapsto D$ is not an FO-product $P\times_{\sigma}P$ because of the
condition $x=y\wedge x^{\prime}=y^{\prime}$ whose equalities concern both
copies of $P$.\ \ \hfill$\square$

\bigskip

\textbf{Theorem 6.3\ (}\emph{Backwards Translation Theorem for vectorial
transductions}): Let be given $\mathcal{R}$, $\mathcal{R}^{\prime}$ and $\tau$
of dimension $k$ as in Definition 6.1. For every FO sentence $\varphi$ over
$\mathcal{R}^{\prime}$, one can construct an FO sentence $\psi$ over
$\mathcal{R}$ such that for every $\mathcal{R}$ structure $S$, we have
$\tau(S)$ $\models$ $\varphi$ if and only if $S\models\psi.$

\textbf{Proof : }Inductive construction based on $\varphi$.\ A subformula
$\theta$\ with free variables is transformed into $\widehat{\theta}.$ The
existential construction $\exists\overline{z}.\theta(\overline{z},...)$ (where
$\overline{z}\in D_{S}^{k}$) yields $\exists z_{1},...,z_{k}.\widehat{\theta
}(z_{1},...,z_{k},...).$\ Because of this step, the quantifier-height
increases.\ We omit the routine details.\ \ \hfill$\square$

\bigskip

\textbf{Proposition 6.4\ }: (1) For every FO (resp. QF)-product $\times
_{\sigma}$ of type $\mathcal{R}\times\mathcal{R}\longrightarrow\mathcal{R}%
^{\prime}$, the transduction $S\longmapsto S\times_{\sigma}S$ \ is a vectorial
FO (resp. QF)- transduction of dimension 2.

(2) For every FO (resp. QF)-product $\times_{\sigma}$ of type $\mathcal{R}%
\times\mathcal{R}^{\prime}\longrightarrow\mathcal{R}^{\prime\prime}$, the
transduction $S\oplus_{m}T\longmapsto S\times_{\sigma}T$ \ is a vectorial FO
(resp. QF)- transduction of dimension 2, with a precondition.

\textbf{Proof} : (1) Clear from definitions.

(2) The domain of $S\times_{\sigma}T$ is the set of pairs $(x,y)$ in
$D_{S}\uplus D_{T}$\ such that $S\oplus_{m}T\models In_{1}(x)\wedge
In_{2}(y).$ The detailed construction is clear.\ \ \hfill$\square$

\section{Recognizability}

We review definitions from \cite{CE},\ Chapters 5 and 7.\ We only consider
finite sets of operations.

\bigskip

\textbf{Definition 7.1\ }: \emph{Recognizability}.

Let $F$ be a\ finite set of operations on a\ set $D$.\ Then $M:=(D,F)$ is an
$F$-\emph{algebra}.\ A subset $L$ of $D$ is $F$-\emph{recognizable} if it is a
union of equivalence classes of a \emph{finite} $F$-\emph{congruence} $\sim$
on $D$.\ This means that $f(d_{1},...,d_{k})\sim f(d_{1}^{\prime}%
,...,d_{k}^{\prime})$ if $d_{i}\sim d_{i}^{\prime}$ for each $i$ and each
operation $f$ of arity $k,$ and that $\sim$\ has finitely many classes.

To distinguish semantics from syntax, we will denote by $f_{M}$ the operation
of $M$\ denoted by the symbol $f$ in $F$.\ The set of terms over $F $ is
$T(F).$ Each term $t$ denotes a finite $\mathcal{R}$-structure $val(t) $. We
say that $M$ is \emph{generated} by $F$ if each element of $D$ is the value
$val(t)$ of some term\ $t$ in $T(F)$; nullary symbols are needed in $F$\ for
having $D$\ not empty. Then, if $L$ is recognizable, the set of terms
$val^{-1}(L)$ is recognized by a finite automaton on $T(F)$.\ Algorithmic
applications are developped in \cite{CE},\ chapter 6.\ \hfill$\square$

\bigskip

We will consider logically defined operations on $\mathcal{R}$-structures,
either unary or binary\footnote{Extension to larger arities is
straightforward.}, in particular those defined in Sections 2,4,5 and 6.\ They
transform concrete structures but preserve isomorphism. Hence, they apply to
\emph{abstract structures}, \emph{i.e.}, to isomorphism classes of
structures.\ If we also have a nullary symbol $\ast$\ denoting a trivial
$\mathcal{R}$-structure consisting of a single element and empty relations,
then the terms generate finite $\mathcal{R}$-structures.

\bigskip

\textbf{Definition 7.2} : \emph{Smooth operations on }$\mathcal{R}%
$-\emph{structures} \emph{and recognizability.}

Let $\mathcal{L}$ be a logical language, say FO\footnote{QF\ sentences are
trivial, either $true$ or $false$, hence are not interesting.} or MSO\ or some
counting extension of them (cf.\ Section 8).\ Let $\varphi$ be a sentence over
$\mathcal{R}$ in $\mathcal{L}$ and $Mod(\varphi)$ be the set of its finite models.

For every positive integer $q$, we let $\sim_{q}$ be the equivalence relation
on $\mathcal{R}$-structures such that $S\sim_{q}T$ if and only if $S$ and $T$
satisfy the same sentences of $\mathcal{L}$ of quantifier-height at most $q$.
Since $\mathcal{R}$ is finite, there are finitely many sentences in
$\mathcal{L}$ of quantifier-height at most $q$, up to logical equivalence
(because, for a typical exampe, $\varphi$\ is equivalent to $\varphi
\wedge...\wedge\varphi$).\ See \cite{CE}, Section 5.6.2, where the cardinality
of the set of equivalence classes is\ effectively bounded.\ It follows that
the equivalence $\sim_{q}$\ has finitely many classes (however, a huge number!)

We say that a binary operation $f$ on $\mathcal{R}$-structures is
$\mathcal{L}$-\emph{smooth}\footnote{A notion introduced by Makowsky
\cite{Mak}.} if for all integer $q$, all $\mathcal{R}$-structures
$S,S^{\prime}$,$T$ and $T^{\prime}$ such that $S\sim_{q}S^{\prime}$ and
$T\sim_{q}T^{\prime}$, then we have $f(S,T)\sim_{q}f(S^{\prime},T^{\prime})$,
and similarly for operations of other arities. It follows that, if the
operations of $F$\ are $\mathcal{L}$-smooth, then each set $Mod(\varphi)$ is
$F$-recognizable where $\varphi$ is a sentence in $\mathcal{L}$, because if
$q$ is the quantifier-height of $\varphi$, then the equivalence $\sim_{q}$\ is
a finite congruence and $Mod(\varphi)$ is a union of classes of it.

\bigskip

\textbf{Theorem 7.3} : The following operations on $\mathcal{R}$ structures
are FO-smooth:

(1) all QF-transductions, either scalar, linearly expanding or vectorial;

(2) disjoint union, either marked or not, and all QF-products.

\textbf{Proof }: Immediate consequences of the Backwards Translation Theorems
(Theorems 2.3, 4.2, 6.3) and the Splitting Theorems (Theorems 5.2,
5.10).\ \ \hfill$\square$

\bigskip

\textbf{Remark 7.4}\ : (1) Up to logical equivalence, there are only finitely
many QF formulas having free variables in a fixed finite set.\ There are thus
finitely operations as above if vectorial transductions have bounded dimension.\ 

(2) For MSO and CMSO, the smooth operations are only the scalar and linearly
expanding QF\ transductions and disjoint union \cite{CE}, Theorem 5.4.7, and
Section 5.3.1.

\bigskip

\textbf{Theorem} \textbf{7.5}\ : If $F$ is a finite set of operations among
those of Theorem 7.3, and $\varphi$ is an FO\ sentence, then the set of terms
$val^{-1}(Mod(\varphi))$ is recognizable in $T(F)$ and thus, definable by a
finite automaton.

\textbf{Proof} : Clear from definitions and Theorem 7.3.\ \ \hfill$\square$

\bigskip

One can add to the operations of Theorem 7.3, the parallel composition of
structures with sources // by Proposition 5.7 and \cite{CE},Theorems 5.47 and 5.57.

\bigskip

The restrictions in Theorem 7.3\ to QF-transductions as opposed to FO-ones are
important, as shown by the following proposition. We need some definitions. We
let $\{a,b\}^{+}$ be the set of nonempty words over letters $a$ and $b$.\ Such
a word $w$ in $\{a,b\}^{+}$ is represented by the nonempty relational
$\mathcal{R}$-structure$\ \lfloor w\rfloor:=(N,\longrightarrow,$
\ $lab_{a},lab_{b})$\ where $N$ is the set of positions, $\longrightarrow$
denotes the successor relation on positions, $lab_{a}(x)$ holds if $x$ is an
occurrence of $a$, and similarly for $b$.

\bigskip

\textbf{Proposition 7.6}: There is a set $H$\ of operations on $\mathcal{\{}%
\longrightarrow,lab_{a},lab_{b},In_{1},In_{2}\mathcal{\}}$-structures
consisting of $\oplus_{m}$, two scalar FO-transductions and two nullary
symbols denoting singleton structures, and, an FO-sentence $\varphi$\ such
that the set of terms $val^{-1}(Mod(\varphi))\subseteq T(H)$ is not
recognizable. The sentence $\varphi$ and the defining formulas of the two
transductions have quantifier-height at most 1.

\textbf{Proof : }We define the following unary operations on $\{a,b\}^{+}$:

\begin{quote}
$f_{a}:u\longmapsto au,$ $f_{b}:u\longmapsto bu$ and $g$ such that $g(u)$ is
obtained from $u$ by replacing simultaneous all factors $ab$ by $a$.\ 
\end{quote}

Hence, $g(aabbaabba)=aababa$, and $g^{n}(aabbaabba)=aaaa$ for every $n\geq2. $

Furthermore $\boldsymbol{a}$ is a nullary symbol\ that denotes the word $a$,
and $\boldsymbol{b}$ for $b$.\ A term of the form $g^{n}(f_{a}(f_{b}%
^{m}(\boldsymbol{b})))$ defines a word without $b$ if and only if $n\geq m+1.$
Hence, by a standard argument on finite automata, these terms do not form a
recognizable set.\ 

It is straightforward to prove that $g$ is a scalar FO-transduction \emph{via}
the representation of a word $w$ by $\left\lfloor w\right\rfloor $.
Furthermore,$\ \lfloor f_{a}(u)\rfloor=\tau(\lfloor\boldsymbol{a}\rfloor
\oplus_{m}\lfloor u\rfloor)$ and$\ \lfloor f_{b}(u)\rfloor=\tau(\lfloor
\boldsymbol{b}\rfloor\oplus_{m}\lfloor u\rfloor)$ for a scalar FO-transduction
$\tau$. Then we let $\varphi$ express that the considered structure has no
element satisfying $lab_{b}$. The sentence $\varphi$ and the formulas defining
$g$ and $\tau$\ have quantifier-height at most 1.

\qquad\hfill$\square$

\section{Counting first-order logic}

We develop a notion introduced\ above in Definition 1.5 and studied already in
\cite{KuSch,NesPOM+}.

\bigskip

\textbf{Definition 8.1\ : }\emph{Modulo counting set predicates.}

For a finite set $A$, the\ set predicate $\mathbb{C}^{(p,q)}(A)\ $means that
$\left\vert A\right\vert =p$ modulo $q,$ where \ $0\leq p<q$ and $q\geq2$. The
following is clear.

\bigskip

\textbf{Lemma 8.2} : Let $A$ and $B$ be disjoint finite sets.\ 

\begin{quote}
(1) $\mathbb{C}^{(p,q)}(A\uplus B)$ $\Longleftrightarrow\bigsqcup$
$\mathbb{C}^{(r,q)}(A)\wedge\mathbb{C}^{(r^{\prime},q)}(B)$
\end{quote}

where the (exclusive) disjunction extends to all $r,r^{\prime}$ in $[0,q-1]$
such that $r+r^{\prime}=p$ mod $q$.\ 

\begin{quote}
(2)\ $\mathbb{C}^{(p,q)}(A\times B)$ $\Longleftrightarrow\bigsqcup$
$\mathbb{C}^{(r,q)}(A)\wedge\mathbb{C}^{(r^{\prime},q)}(B)$
\end{quote}

where the disjunction extends to all $r,r^{\prime}$ in $[0,q-1]$ such that
$r\times r^{\prime}=p$ mod $q$.\ \ \ \hfill$\square$

\bigskip

These equalities extend to combinations of Cartesian products and disjoint
unions of several sets.\ For example, if $A:=B\uplus C\times(D\uplus(E\times
F))$ then \ 

\begin{quote}
$\mathbb{C}^{(p,q)}(A)$ $\Longleftrightarrow$ $\bigsqcup$ $\mathbb{C}%
^{(b,q)}(B)\wedge\mathbb{C}^{(c,q)}(C)\wedge\mathbb{C}^{(d,q)}(D)\wedge
\mathbb{C}^{(e,q)}(E)\wedge\mathbb{C}^{(f,q)}(F)$
\end{quote}

where the exclusive disjunction extends to all integers $b,c,d,e,f$ in
$[0,q-1]$ such that $b+c(d+e\times f))=p$ mod $q$.

\bigskip

\textbf{Definition 8.3\ : }\emph{Counting First-Order logic (C}$^{+}%
$\emph{FO)}

For defining \emph{C}$^{+}$\emph{FO}\ formulas, we use the \emph{modulo
counting} construction $\mathbb{C}^{(p,q)}\overline{x}.\varphi(\overline
{x},\overline{y}$) where $\overline{x}$ \ and $\overline{y}$ are disjoint
tuples of free variables and $\varphi$ is C$^{+}$FO.\ It yields a formula
$\psi$ with free variables in $\overline{y}$.\ Then $(S,\overline{b}%
)\models\psi$ (where $\overline{b}$ defines values for $\overline{y}$) if the
set of tuples $\overline{a}$ such that $(S,\overline{a},\overline{b}%
)\models\varphi$ has cardinality $p$ modulo $q$.

We refer by CFO to formulas that do not count tuples but only individual
elements.\ Note that $\mathbb{C}^{(p,q)}x.\varphi(x,\overline{y}$) can be
expressed by a \emph{monadic second-order} formula written with the modulo
counting set predicate $\mathbb{C}^{(p,q)}(X)$, hence by a formula of CMSO
logic (cf.\ \cite{CE}, Definition 5.24).

At the cost of adding existential quantifications, one can replace the
generalized quantifiers $\mathbb{C}^{(p,q)}\overline{x}$ for $p>0$ by
$\mathbb{C}^{(0,q)}\overline{x}$.\ Furthermore, the next theorem proves that
counting tuples, as opposed to individual elements, brings no additional
expressive power.

\bigskip

\textbf{Theorem 8.4\ } : Every formula $\mathbb{C}^{(p,q)}\overline{x}%
.\varphi$ where $\varphi$ is\ C$^{+}$FO\ and $\overline{x}=(x_{1},...,x_{m})$
can be expressed by a CFO\ formula using only the constructions $\mathbb{C}%
^{(p^{\prime},q)}x_{i}$.$\varphi^{\prime}$ for $0\leq p^{\prime}<q$ and
CFO\ formulas $\varphi^{\prime}$.

\textbf{Proof : }We first consider the case of simple directed graphs
(possibly with loops).\ 

The formula $\mathbb{C}^{(0,2)}(x,y).edg(x,y)$ expresses that such a graph has
an even number of arcs. The formula $\mathbb{C}^{(0,2)}x.edg(x,x)$ expresses
that it has an even number of loops. The formula $\mathbb{C}^{(0,2)}%
x.\mathbb{C}^{(1,2)}y.(x\neq y\wedge edg(x,y))$ expresses that it has an even
number of non-loop arcs: it says that the number of vertices $x$ of odd
outgoing degree (excluding loops) is even.\ By combining these formulas and
their negations, one gets the desired formula equivalent to $\mathbb{C}%
^{(0,2)}(x,y).edg(x,y)$.

The same idea can be used for $\mathbb{C}^{(p,q)}(x,y).edg(x,y)$ and any
$0\leq p<q,$ whence also for any formula $\mathbb{C}^{(p,q)}(x,y).\varphi
(x,y)$ by replacing $edg(x,y)$ by $\varphi(x,y)$.

We now consider the case of $\mathbb{C}^{(p,q)}\overline{u}.\varphi$.\ The
proof is by induction on $m$.

We consider $\overline{x}$ as the pair $(x_{1},\overline{w})$ where
$\overline{w}=(x_{2},...,x_{m}).$ We apply the previous construction with
$x:=x_{1}$ and $y:=\overline{w}$.\ It remains to handle formulas of the form
$\mathbb{C}^{(p^{\prime},q)}\overline{w}.\varphi^{\prime}$.\ We are reduced to
the case $m-1$, and we use induction.\ We omit further details. \ \ \hfill
$\square$

\bigskip

The article \cite{KuSch} considers CFO\ formulas built with $\mathbb{C}%
^{(p,q)}x$ as opposed to $\mathbb{C}^{(p,q)}\overline{x}$ and establishes a
relevant extension of Gaifman's Locality Theorem.

\bigskip

\textbf{Question 8.5\ }: For how much should we count $\mathbb{C}%
^{(p,q)}\overline{x}\ \ $in evaluating quantifier-heights?

We have seen in Section 7\ that quantifier-height is important for proving
recognizability because, up to logical equivalence, there are finitely many
FO\ (and even MSO) sentences of quantifier-height at most $h$ (over a fixed
signature) for any given $h$.\ See \cite{CE}\ Propositions 5.93 and
5.94.\ This yields the Recognizability Theorem for FO logic of Theorem 7.5.\ 

There are infinitely many inequivalent sentences of the form

\begin{quote}
$\mathbb{C}^{(p,q)}\overline{x}.R(x_{1},x_{2})\wedge R(x_{2},x_{3}%
)\wedge.....\wedge R(x_{k-1},x_{k})$
\end{quote}

where $\overline{x}=(x_{1},...,x_{k})$ and $k$ is unbounded. Hence we cannot
define as 0 or 1 the contribution of $\mathbb{C}^{(p,q)}\overline{x}$ to
quantifier-height to obtain the finiteness result yielding Theorem 7.5. We
will count it for $k$.\ This is coherent with Theorem 8.4.\ The proofs of
Propositions 5.93 and 5.94 of \cite{CE}\ \ extend easily.\ 

Note that in \emph{Counting Monadic Second-Order} (CMSO) logic, $\mathbb{C}%
^{(p,q)}(X)$ is a set predicate of quantifier-height 0, and not a generalized
quantifier, whereas in C$^{+}$FO, $\mathbb{C}^{(p,q)}(x,y).R(x,y) $ is not quantifier-free.

\bigskip

The following theorem generalizes Theorems 2.3 and 4.2.

\bigskip

\textbf{Theorem 8.6} (\emph{Backwards Translation Theorem}).

(1) Let $\tau$ be a linear expanding CFO-transduction without parameters of
type $\mathcal{R}\rightarrow\mathcal{R}^{\prime}$.\ Let $\varphi$ be a CFO
sentence of signature $\mathcal{R}^{\prime}$.\ There exists a CFO sentence
$\psi$\ such that, for all $\mathcal{R}$-structure $S$ and $\mathcal{R}%
^{\prime}$-structure $T$ such that $T=\tau(S)$, then we have:

\begin{quote}
$T\models\mathbb{\varphi}$ $\Longleftrightarrow$ $S\models\mathbb{\psi}$ .
\end{quote}

(2) If $\tau$ is QF, then the quantifier-height of $\psi$ is no larger than
that of $\varphi$.

(3) Property (1) holds if $\tau$\ is vectorial.

\textbf{Proof}: (1) We first consider the\ case of a scalar transduction, and
we adapt the construction of Theorem 2.3.\ We need to perform the induction
for formulas having free variables. Consider $\varphi(x_{1},...,x_{n})$, an
$(n-1)$-tuple $\overline{a}$ in $D_{S}^{n-1}$ and $b$ in $D_{S}.$

Then, by Theorem 2.3(1), we have, for some formula $\psi$

\begin{quote}
$(T,\overline{a}b)\models\varphi$ $\Longleftrightarrow$ $(S,\overline
{a}b)\models\delta(x_{1})\wedge...\wedge\delta(x_{n})\wedge\psi(x_{1}%
,...,x_{n}).$
\end{quote}

Hence,

$\qquad(T,\overline{a})\models\mathbb{C}^{(p,q)}x_{n}.\varphi(x_{1}%
,...,x_{n})$ $\Longleftrightarrow$

\qquad$(S,\overline{a})\models$\ $\delta(x_{1})\wedge...\wedge\delta
(x_{n-1})\wedge\mathbb{C}^{(p,q)}x_{n}.[\delta(x_{n})\wedge\psi(x_{1}%
,...,x_{n})].$

Note that the definitions of the relations in $T$ (hence the formulas $\psi$)
can use the generalized quantifier $\mathbb{C}^{(p,q)}$\ since $\tau$ may be a CFO-transduction.

If $\tau$ is QF\ then the quantifier-height of $\psi$ is no larger than that
of $\varphi$, this is clear from the inductive step.

\bigskip

Next we consider the case of a $k$-copying linearly expanding
CFO-transduction. Here $D_{T}=D_{1}\times\{1\}\cup...\cup D_{k}\times\{k\} $
where each subset $D_{j}$ of $D_{S}$ is defined in $S$ by some formula
$\delta_{j}(x)$.\ 

The set $B_{\overline{a}}$ of elements $b$ such that $(T,\overline{a}%
b)\models\varphi$ is the disjoint union of $k$ sets $B_{\overline{a},j}%
\times\{j\}$ where $B_{\overline{a},j}\subseteq D_{j}$ is defined by a CFO
formula. The objective is\ to apply Lemma 8.2(1).\ 

A complication comes from the fact that the components of $\overline{a}$
belong to $k$ different subsets $D_{j}\times\{j\}$ of $D_{T}$.\ There are thus
$k^{n-1}$ types of tuples.\ We take $n=2$\ to simplify the exposition.\ The
general case is just more difficult to write down.

Our aim is to express $\mathbb{C}^{(p,q)}x_{2}.\varphi(x_{1},x_{2})$ in $S$.\ 

We define the formula $\varphi_{i,j}(x_{1},x_{2})$ over $\mathcal{R}^{\prime
}\cup\{In_{1},...,In_{k}\}$ as $In_{i}(x_{1})\wedge In_{j}(x_{2})\wedge
\varphi(x_{1},x_{2}).$

By arguing as in the proof of Theorem 7.10\ of \cite{CE}, one can construct
formulas $\psi_{i,j}(x_{1},x_{2})$ over $\mathcal{R}$ from the $\varphi_{i,j}%
$'s, such that, for any $a^{\prime}$ and $b^{\prime}$ in $D_{S}$,

\begin{quote}
$T\models\varphi((a^{\prime},i),(b^{\prime},j))$ $\Longleftrightarrow$
$S\models\delta_{i}(a^{\prime})\wedge\delta_{j}(b^{\prime})\wedge\psi
_{i,j}(a^{\prime},b^{\prime}).$
\end{quote}

Hence, if we let $a=(a^{\prime},i)$ with $S\models\delta_{i}(a^{\prime})$ then
$B_{a,j}$ as defined above is the set of $b^{\prime}$ such that $S\models
\delta_{j}(b^{\prime})\wedge\psi_{i,j}(a^{\prime},b^{\prime})$ for some $j$.
By using Lemma 8.2(1), one can express $\mathbb{C}^{(p,q)}(B_{a,i})$ as a
Boolean combination of the formulas $\mathbb{C}^{(p^{\prime},q)}x_{2}%
.(\delta_{j}(x_{2})\wedge\psi_{i,j}(a^{\prime},x_{2}))$ for the $k$ different
$j$'s and fixed $i$. We obtain a CFO\ formula $\theta_{i}(x_{1})$ such that
for all $a^{\prime}$ in $D_{S}$\ :

$S\models\theta_{i}(a^{\prime})$ $\Longleftrightarrow$ $\mathbb{C}%
^{(p,q)}(B_{a,i})$ is true where $a=(a^{\prime},i)$.

Then for each $a$ in $D_{T}$ with first component $a^{\prime}$:

$(T,a)\models\mathbb{C}^{(p,q)}x_{2}.\varphi(x_{1},x_{2})$ if and only
if\ ($S,a)\models$ $\bigvee\nolimits_{i\in\lbrack k]}\delta_{i}(x_{1}%
)\wedge\theta_{i}(x_{1})$.

We can convert this last formula into an exclusive disjunction, and use then
Lemma 8.2(1).\ 

(3) Let $\tau$ be vectorial of dimension $k$, and $T=\tau(S)$. We expand the
proof of Theorem 6.3. We must express in $S$ the validity in $T$ of the
formula $\mathbb{C}^{(p,q)}\overline{x}_{n}.\varphi(\overline{x}%
_{1},...,\overline{x}_{n})$ where the variables $\overline{x}_{1}%
,...,\overline{x}_{n} $ ranging over $D_{T}$\ are actally $k$-tuples of
elements of $D_{S}$. Again we take $n=2$.

Hence, $(T,\overline{a}_{1})\models\mathbb{C}^{(p,q)}\overline{x}_{2}%
.\varphi(\overline{x}_{1},\overline{x}_{2})$ \ if and only if $(S,\overline
{a}_{1})\models$ $\mathbb{C}^{(p,q)}\overline{x}_{2}.\widehat{\varphi
}(\overline{x}_{1},\overline{x}_{2})$ (we use the construction of Theorem
6.3).\ As $\overline{x}_{2}$ is a $k$-tuple of variables, we have a C$^{+}$FO
formula.\ By Theorem 8.4, it can be transformed into a CFO formula.\ 

\ 

\textbf{Theorem 8.7} (\emph{Splitting Theorem for binary operations).}

(1) Theorem 5.4 holds for $\oplus$\ and $\oplus_{m}$\ and CFO formulas.

(2) Theorem 5.13 holds for a CFO product $\otimes_{\sigma}$\ and CFO formulas.

\textbf{Proof} : (1) \ This case of is proved in Theorem 3.1\ of \cite{KuSch}.

(2) We consider a CFO\ product of structures $S\otimes_{\sigma}T$.\ 

We adapt the proof of Lemma 5.12(1) and we conclude by using Theorem 8.4 in
order to \ \ 

to obtain a CFO formula.\ 

{\Large Conclusion}

Our objective was to clarify and compare different types of first-order
transductions of relational structures, and to relate them to binary operations.\ 

Except for recognizability, our definitions and results extend to countable
structures where $\mathbb{C}^{(p,q)}x.\varphi(x,\overline{y})$ means that the
number of elements $x$ that satisfy $\varphi(x,\overline{y})$\ \ is finite and
of cardinality $p$ modulo $q$.

\end{document}